\documentclass[sn-mathphys-num]{sn-jnl}% Math and Physical Sciences Numbered Reference Style 
\usepackage{graphicx}%
\usepackage{multirow}%
\usepackage{amsmath,amssymb,amsfonts}%
\usepackage{amsthm}%
\usepackage{mathrsfs}%
\usepackage[title]{appendix}%
\usepackage{textcomp}%
\usepackage{manyfoot}%
\usepackage{booktabs}%
\usepackage{algorithm}%
\usepackage{algorithmicx}%
\usepackage{algpseudocode}%
\usepackage{listings}%
\usepackage{caption}
\usepackage{subcaption}
\usepackage[table]{xcolor}
\usepackage{enumitem}
\theoremstyle{thmstyleone}%
\theoremstyle{thmstyletwo}%

\theoremstyle{thmstylethree}%

\newcommand{\RomanNumeralCaps}[1]
    {\MakeUppercase{\romannumeral #1}}

\newcommand{\revision}[1]{#1}  %revision not shown

\newcommand\upstrut{\rule{0pt}{7.5pt}}
\newcommand\downstrut{\rule[-6pt]{0pt}{6pt}}
\newcommand\mystrut{\upstrut\downstrut}

\begin{document}

\title[Augmenting the action space with conventions to improve multi-agent cooperation in Hanabi]{Augmenting the action space with conventions to improve multi-agent cooperation in Hanabi}


\author*[1]{\fnm{F.} \sur{Bredell}}\email{20795718@sun.ac.za}

\author[1]{\fnm{H. A.} \sur{Engelbrecht}}\email{hebrecht@sun.ac.za}

\author[1]{\fnm{J. C.} \sur{Schoeman}}\email{jcschoeman@sun.ac.za}

\affil*[1]{\orgdiv{Electrical and Electronic Engineering}, \orgname{University of Stellenbosch}, \state{Western Cape}, \country{South Africa}}

\abstract{ 
    The card game \textit{Hanabi} is considered a strong medium for the testing and development of multi-agent reinforcement learning (MARL) algorithms, due to its cooperative
    nature, \revision{partial observability}, limited communication and remarkable complexity. Previous research efforts have explored the capabilities of MARL algorithms within
    Hanabi, focusing largely on advanced architecture design and algorithmic manipulations to achieve state-of-the-art performance for various number of cooperators. However, this
    often leads to complex solution strategies with high computational cost and requiring large amounts of training data. For humans to solve the Hanabi game effectively, they
    require the use of conventions, which often allows for a means to implicitly convey ideas or knowledge based on a predefined, and mutually agreed upon, set of ``rules'' or
    principles. Multi-agent problems containing partial observability, especially when limited communication is present, can benefit greatly from the use of implicit knowledge
    sharing. In this paper, we propose a novel approach to augmenting an agent's action space using \emph{conventions}, which act as a sequence of special cooperative actions that
    span over and include multiple time steps and multiple agents, requiring agents to actively opt in for it to reach fruition. These \emph{conventions} are based on existing
    human conventions, and result in a significant improvement on the performance of existing techniques for self-play and cross-play \revision{for} various number of cooperators
    within Hanabi. }

\keywords{Multi-agent Reinforcement Learning, Hanabi, Agent Cooperation, \revision{Self-play}, Cross-play}



\maketitle

\section{Introduction}\label{intro} 
    Reinforcement learning (RL) holds promising potential to address a large variety of problems where artificial agent operation offers a significant improvement over alternative
    methods, such as hard-coded algorithms or solutions~\cite{marl_why_good}. For certain real-world problems, single-agent operation is not optimal (or even possible) and the
    incorporation of multiple agents would be beneficial (or necessary). A prime example is autonomous vehicles navigating roads~\cite{automated_MARL_vehicles}, where each vehicle
    is controlled by an autonomous agent and these agents must cooperate effectively to ensure road safety. Other examples of problems that benefit from multiple agents include
    guided drone swarms~\cite{guided_drone_swarms}, mapping verbal instructions to executable actions~\cite{map_instr_to_act}, or the switching of railway lines. RL can help agents
    to effectively cooperate within these multi-agent scenarios by learning from past and or simulated experiences. 

    Unfortunately, the introduction of multiple agents into an environment typically increases the complexity of the problem exponentially. It introduces a moving target learning
    problem~\cite{MARL_survey}, since all the agents must learn simultaneously. The reason for this is that each individual agent's policy changes over time, which in turn causes a
    non-stationary environment. This often inhibits all the agents from developing effective policies and can lead to undesired behaviour~\cite{MARL_survey}. Furthermore,
    multi-agent systems often contain partial observability, where the full state space is hidden from the individual agents, resulting in each agent having their own unique
    perspective of the problem. In the case of autonomous agents controlling vehicles, even though an agent might have access to high-definition sensors (often used in vehicle
    control), the environment will still contain objects outside the current agent's perspective, for example objects outside the sensor's range or viewing angle or objects
    obscured by other vehicles, static objects, blind corners, concealed driveways, etc. 
    
    When another agent controlling a different vehicle is introduced into the environment, the need for communication to overcome the problem of partial observability becomes
    apparent, since this agent will have a different viewpoint of the environment. These viewpoints will have some overlap, but will also contain important information that the
    agents will have to share, such as obscured objects, accidents, obstacles, pedestrians, etc., in order to safely navigate the road. Even when the agents are able to
    communicate, it is infeasible to share and process all the \revision{high-quality} information between agents, especially in scenarios where quick reaction times are crucial.
    This communication problem becomes even more difficult if we consider situations where more than two vehicles must coordinate effectively, for example a large intersection,
    road construction, an accident that lead to congestion, or a busy city. 
    
    One of the most straightforward solutions to multi-agent reinforcement learning (MARL) is the combination of deep Q-networks (DQNs) and independent Q-learning~\cite{idqn_pong},
    where each agent independently and simultaneously learns its own action-value function (or \textit{Q-values}\footnote{Action values ($Q$) refer to the value assigned to a
    certain action $a$ within a given state $s$.}) while interacting with the same environment. However, this strategy does not fair well when paired with partial
    observability~\cite{iql_short_og, iql_short_new}. Hausknecht and Stone~\cite{drqn_stone} have shown that recurrent neural networks offer an improved solution to MARL problems
    containing partial observability. These networks incorporate a built-in \revision{short-term} memory over which experiences are \emph{unrolled} (or combined) to form longer
    sequences. This is often combined with deep Q-learning's feed forward neural network to produce deep recurrent Q-learning~\cite{drqn_stone}. 

    Various research efforts follow a similar path as Hausknecht and Stone, focusing on architectural advancements through the means of neural network layer manipulations, and
    complex algorithms to improve the estimation and assignment of action-values or policies~\cite{vdns, rashid2018qmix, foerster_rial_dial}. This often leads to convoluted
    solution strategies that are difficult to implement and computationally expensive~\cite{sad_hanabi, sad_search}. An alternative solution strategy would be to reconsider the
    problem dynamics and discover ways of incorporating existing domain knowledge into RL algorithms. Existing domain knowledge can be incorporated into RL and MARL using reward
    shaping~\cite{og_reward_shaping_1,og_reward_shaping_2,marl_reward_shaping}, which focuses on manipulating the reward signal to encourage certain behaviour. Alternatively,
    agents can use state-space augmentations, such as auxiliary tasks~\cite{aux_tasks,aux_tasks_2}, to imply or extract additional information (or features) of the problem setting.
    \revision{Sutton} \textit{et al.}~\cite{sutton1999options} have shown that \textit{options} offer an additional strategy for incorporating domain knowledge into RL by changing
    the action space of an agent. Options allow an agent to solve problems on a higher level by extending the action space to include advanced temporal actions that range over
    multiple time steps (also referred to as \textit{macro-actions}). 

    Games are often used as a medium for testing and evaluating RL algorithms, since they incorporate real-world problems with well-defined rules and a clear metric for measuring
    performance, where examples include \textit{Go}~\cite{silver2016alphago}, \textit{Backgammon}~\cite{gammon}, and \textit{Dota 2}~\cite{openai_five}. \textit{Hanabi} is a
    cooperative card game containing partial observability and limited communication, requiring players to logically reason over the intentions and actions of their cooperators, a
    concept known as theory of mind~\cite{ToM}. Human players require the development of conventions\footnote{Philosopher David Lewis defines conventions as an arbitrary,
    self-perpetuating solution to a recurring coordination problem~\cite{lewis2008convention}.} in order to solve the problem effectively. Conventions have evolved with humanity
    throughout the \revision{ages} and range from driving on a certain side of the road to social conventions or norms~\cite{lewis2008convention}. 

    Conventions allow for additional implicit communication through actions and mutually agreed upon ``rules'', thereby overcoming the communication restrictions of the
    communication channel. Similar to the Hanabi problem, when humans tackle the problem of controlling vehicles on roads, conventions are common practice and ensure effective
    cooperation. Examples include: driving on a certain side of the road\footnote{Philosopher David Lewis discusses this topic in depth, and places particular focus on how this
    affects other aspects of life, such as people tending to keep to the same side as their road conventions when walking in a busy street~\cite{lewis2008convention}.}; keep left,
    pass right (or vice versa); who has \revision{right} of passage at a T-intersection or four-way intersection (and how this changes based on the presence of pedestrians);
    flashing of headlights while stationary to indicate giving right of way/yielding; flashing of headlights while moving to indicate danger ahead, or be wary of potential danger
    ahead; flashing of hazards to indicate an emergency; a quick flash of hazards to say thank you; and many more. Furthermore, this highlights the importance \revision{of
    artificial agents learning existing human conventions}, since there will \revision{come a time} when autonomous agents must navigate these complex scenarios alongside humans,
    and the agents will have to be able to convey and receive information effectively without using their agent-specific communication channel. 
    
    Most research efforts on the Hanabi problem use complex architecture design with advanced algorithms, often focusing on estimating and calculating the
    \revision{\textit{beliefs}} of other agents~\cite{bad,sad_hanabi}. However, one avenue of the Hanabi problem yet to be explored is that of conventions, and how to incorporate
    existing human conventions into MARL solution strategies. In this paper, we propose a method to incorporate human conventions into MARL through the use of artificial
    \emph{conventions}, which \revision{act} as cooperative actions that span over and include multiple time steps and multiple agents, \revision{as well as} show how it
    significantly improves on the performance of MARL agents within Hanabi for self-play~\cite{self_play} as well as cross-play~\cite{otherplay} scenarios. Our approach shares
    similarities with options, due to the multi-time step extension of actions, however is fundamentally built on a different concept. 

\subsection{Related Work}\label{related}  
    A popular solution strategy to RL problems introduced by Hessel \textit{et al.}~\cite{rainbow}, called \textit{Rainbow}, combines various advancements made to deep Q-learning
    and has proven to offer significant performance gain when faced with large discrete action spaces~\cite{rainbow_energy,rainbow_good_1,rainbow_good_2}. A natural extension of
    Rainbow to multi-agent systems, is the implementation of independent Rainbow agents, referred to as multi-agent Rainbow (MA-Rainbow)~\cite{marainbow}. However, Rainbow
    introduces a plethora of new hyperparameters which can often result in suboptimal policy development for cooperative settings~\cite{rainbow_shortcoming}, and is known for
    falling into suboptimal local minima~\cite{rainbow_suboptimal_1,rainbow_suboptimal_2,hanabi_ai}. 

    Hanabi was first proposed as a viable medium and frontier for MARL by Bard \textit{et al.}~\cite{hanabi_ai}, with focus placed on the philosophical ideas and challenges found
    within the problem setting and how they translate to real-world scenarios. Bard \textit{et al.} conducted initial tests for three different MARL algorithms, namely MA-Rainbow,
    actor-critic-Hanabi-agent (ACHA)~\cite{hanabi_ai}, and Bayesian action decoder (BAD)~\cite{bad}, \revision{comparing} them to state-of-the-art handcrafted bots, such as
    \textit{HatBot}~\cite{hatbot} and \textit{WTFWThat}~\cite{wtfwthat}. In the most difficult scenario of five players, the MARL solutions achieved an acceptable score of 16.8/25,
    however this still significantly falls behind the best handcrafted bot able to achieve a score of 24.89/25.

    The simplified action decoder (SAD), introduced by Hu and Foerster~\cite{sad_hanabi}, aims to solve a similar goal as the BAD algorithm, but with the main benefit of reducing
    the effect exploration actions have on the public beliefs of agents~\cite{sad_hanabi}. They improve on the performance of the BAD algorithm, while also extending these concepts
    to more than just two players. Recurrent neural networks~\cite{drqn_stone} are used to track the public beliefs and an auxiliary task~\cite{aux_tasks} is trained to predict if
    a card is playable, discardable or unknown. Hu and Foerster found that these auxiliary tasks only benefited the \revision{two-player} scenario while drastically hurting the
    3--5 player performance~\cite{sad_hanabi}. They are able to achieve good performance for 2--5 players in Hanabi, and close the gap between RL agents and handcrafted solutions
    by achieving a score of 22.06/25 in \revision{five-player} Hanabi. However, the SAD algorithm is computationally expensive, requiring a minimum of 40 CPU cores, 2 high-end GPUs
    and 256 gigabytes of RAM, while still requiring billions of samples and a wall time of 72 hours~\cite{sad_hanabi}. 

    Lerer \textit{et al.}~\cite{sad_search} further builds on the concept of the SAD algorithm by introducing multi-agent tree search (MATS)~\cite{multi_agent_search}, and manage
    to achieve state-of-the-art performance for 2--5 player Hanabi. The agents manage to beat the current best handcrafted bots in the \revision{two-player} scenario, with the
    handcrafted bots only outperforming the agents in \revision{five-player} by 4\%. However, just like its predecessor, the SAD-MATS algorithm is computationally expensive and
    requires a substantial amount of training data and wall time to achieve these results~\cite{sad_search}. 

    Hu \textit{et al.}~\cite{otherplay} also builds on the concept of the SAD algorithm, but rather than achieving state-of-the-art performance for self-play agents, they focus on
    the agent's ability to cooperate with a variety of partners--a concept known as cross-play. This is achieved using \textit{other-play}, an architecture for multi-agent systems
    specifically designed to break the ties formed between self-play agents, allowing them to produce more robust strategies. This allows the agents to cooperate with agents from
    different training runs (or regimes), and even human cooperators~\cite{otherplay}. Even though the other-play agents are able to achieve effective cross-play performance when
    paired with a variety of agents in a \revision{two-player} scenario, the research does not focus on higher player counts where more advanced cooperative strategies are
    required. Additionally, other-play builds on the existing architecture of the SAD algorithm leading to increased computational complexity and cost~\cite{otherplay}.

    In contrast to value-based methods, Yu \textit{et al.}~\cite{mappo} show that multi-agent proximal policy optimization (MAPPO) offers similar performance to value-based
    alternatives in four different cooperative tasks, namely multi-agent particle world environment (MPE)~\cite{mpe}, StarCraft micromanagement challenge (SMAC)~\cite{smac}, Google
    research football (GRF)~\cite{grf}, and the Hanabi challenge~\cite{hanabi_ai}. MAPPO is able to match, and in some instances outperform the capabilities of leading edge
    value-based methods, such as Qplex~\cite{qplex}, RODE~\cite{rode}, and value decomposition networks (VDNs)~\cite{vdns}. It shows promising results in Hanabi and often
    outperformed MA-Rainbow, but is sample inefficient requiring a substantial amount of training steps (more than 10 billion) to achieve good performance. 
    
\subsection{Summary of Contributions}
    Rather than focusing on complex architecture design, this research shifts focus to incorporating existing domain knowledge of the Hanabi problem into MARL algorithms. Hanabi
    has an active and dedicated community \revision{of} players constantly searching for new conventions and ways to reliably beat the game. These conventions are often based on
    principles\footnote{A convention principle is a strategy, or user defined ``rule'', external to the existing game rules which governs a player's behaviour and results in a
    desired outcome.}, and have generally been standardised within the Hanabi \revision{player-base}. Our main contribution is to show how these existing human conventions can be
    implemented in a MARL scenario using artificial \emph{conventions}, which act as a special form of cooperative actions and can be incorporated into existing MARL algorithms
    using action space augmentation. 

    One of the key insights to our approach is the ``subscribing'' technique for \emph{convention} continuation, where an agent can initiate a \emph{convention} (based on a certain
    human convention) and subsequently a new action appears to the other agents to ``subscribe'' or continue with this \emph{convention}. This strategy requires that agents opt in
    to a certain \emph{convention} in order for it to reach completion, and allows an agent to halt a certain inferior \emph{convention}, and/or initiate an improved
    \emph{convention}, based on their unique observation. To this end, each \emph{convention} contains an initial condition acting as the start of a \emph{convention}, subsequent
    continuation conditions for other agents to opt in, a termination condition for a terminating action, and finally a policy that translates each step of the \emph{convention}
    into an environment action based on the current agent's observation.
    
    Our approach shares similarities with that of multi-agent options, but addresses a problem where agents must often wait for other agents to complete their options before they
    can initiate a new option (which can include a cooperative option)~\cite{macdec_pomdp}. Fundamentally, our approach is built on a different concept, due to the cooperative
    requirement of conventions to reach fruition, and their ability to convey ideas or intentions through implicit communication based on mutually agreed upon ``rules''. Therefore,
    it is important to note that our approach does not introduce additional explicit communication, or additional communication channels for that matter, but rather allows for
    implicit communication through the nature of conventions. Furthermore, each agent forming part of \revision{a} \emph{convention} \revision{must} actively \revision{choose to
    participate} in every step of the \emph{convention}, rather than an external policy taking control and executing over multiple time steps until the option terminates.

    Ultimately, we believe our work will act as the foundation for \textit{convention} discovery, i.e. the ability for agents to define their \emph{conventions} as they train.
    Similar to how options paved the way for option discovery by showing the importance and benefits of options, \textit{conventions} offer a similar argument for multi-agent
    cooperation in partially observable environments. Option discovery has been proven to be a difficult problem to solve in multi-agent
    systems~\cite{macdec_pomdp,option_discovery_marl}, and we believe \textit{convention} discovery would offer a similar challenge. Therefore, it is important to first prove the
    capabilities of conventions, their implications, implementation, and limitations, especially at the hands of a difficult problem setting such as Hanabi.
    
    \emph{Conventions} result in a large performance uplift for existing MARL techniques, specifically multi-agent Rainbow, by significantly decreasing the training time and, in
    the case of 3--5 player Hanabi, increasing the converged performance. Additionally, since the agents are learning from a communal and well-defined list of principles, the
    agents are not just able to cooperate in self-play, but also in a cross-play setting with agents from different training runs (or regimes). Cross-play has been shown to be an
    important research area for MARL, since it opens the door to agents being able to cooperate with never-before-seen teammates, whether they are agents from other regimes, agents
    with different architectures, or humans (such as autonomous agents navigating the road alongside humans)~\cite{sp_vs_cp_1,sp_vs_cp_2}. 

\subsection{Paper Outline}
    In Section~\ref{prelim} we introduce the background and theory of RL, MARL, and options, in order to give context for our approach. This is followed by the introduction and
    discussion of \emph{conventions} and action space augmentation with \emph{conventions} in Section~\ref{conventions}. In Section~\ref{hanabi}, we introduce the Hanabi problem
    and learning environment, as well as existing human conventions and how they translate to \emph{conventions} using our discussed formulations and definitions.
    Section~\ref{hanabi} also contains the results for initial tests conducted on the Small Hanabi environment, to prove the capabilities and benefits of using action space
    augmentation with \emph{conventions}. Finally, we present and discuss our results for self-play and cross-play for all player counts in Hanabi, in Section~\ref{results}. 

\section{Multi-agent Reinforcement Learning and Options}\label{prelim} 
    Reinforcement learning aims at solving decision-making problems and is built on the concept of Markov decision processes (MDPs)~\cite{sutton_barto}. Multi-agent reinforcement
    learning introduces more than one agent into the environment and often incorporates partial observability, and thus is built on the concept of decentralised partially
    observable Markov decision processes (Dec-POMDPs)~\cite{dec_pomdp}. The Dec-POMDP framework consists of
    \begin{equation}\label{dec_pomdp}
        \langle S, A, T, O, P, R, n, \gamma \rangle,
    \end{equation}
    where $S$ denotes the global state space and $A$ represents the joint-action space of the $n$ agents at time $t$. The observation space $O$ consists of the local observation of
    each agent $i$ at time $t$ within the global state $S$ ($O^i_t = \mathcal{O}(S;i;t)$). After $A$ is sent to the environment, the global state transitions from $S$ to $S'$ given
    the state transition probability $T$, and similarly $P$ represents the observation transition probability to transition from $O$ to $O'$. This results in the environment
    producing $R$, which denotes the immediate reward for each agent action that contributed to a global state change at $t$, with $\gamma$ denoting the discount factor for future
    reward. These rewards are either combined with equal weighting (summed), or can have their own discount factor $\gamma_r$, and is referred to as the forward accumulated
    reward~\cite{dec_pomdp}.
    
    In value-based RL we can use \revision{equation} (\ref{dec_pomdp}) to calculate the value functions ($Q$) which acts as a quantitative measure for the desirability of a state.
    In tabular methods these values are represented using a lookup-table, while deep RL methods use powerful function approximators to estimate these values. 

\subsection{Independent Q-learning}\label{iql}
    Q-learning is an off-policy, temporal difference (TD) control algorithm that learns the action-value function $Q(S,A)$ by directly approximating the optimal action-value
    function $Q^*(S,A)$, independent of the policy being followed~\cite{sutton_barto}. The update step for the action-value function is defined as
    \begin{equation}\label{update_step_ql}
        \begin{aligned}
            Q(S_t, A_t) \gets {} & Q(S_t, A_t) + \alpha \bigr[ R_{t+1} + \gamma \max_{a} Q(S_{t+1}, a) - Q(S_t, A_t) \bigr],
        \end{aligned}
    \end{equation}
    where $\alpha$ is the learning rate and $\gamma$ the discount factor of future rewards~\cite{sutton_barto}. This can be extended to MARL with independent Q-learning, where each
    agent has their own action-value function, conditioned on their observations $O^i_t$ instead of the global state $S_t$, and using their individual experiences in their update
    steps~\cite{iql_og}. The algorithm for independent Q-learning in a turn-based scenario is shown in Algorithm~\ref{ql algorithm}, where each agent $i$ uses their own
    observations $O^i_t$ to update their action-value functions $Q^i$. The algorithm receives as input all the hyperparameters for the agent's architecture, and produces a policy
    as output in the form of the action-value function.

    \begin{algorithm}
        \caption{Independent Q-learning in a turn-based environment}\label{ql algorithm}
        \begin{algorithmic}[1]
            \State Initialise hyperparameters: learning rate $\alpha \in (\,0, 1]\,$, discount factor $\gamma \in (\,0, 1]\,$, total number of player $P$, and exploration rate $\epsilon \in (\,0, 1)\,$
            \State Initialise all $Q(O, A) \gets 0$

            \For{each episode}
                \State Reset environment and set total time $t \gets 0$

                \While{$S_{t+1}$ is not terminal}
                    \For{all players $i$ from $0$ to $P-1$}
                        \State $A^i_t = \text{argmax}_a Q^i(O^i_t, a)$ \Comment{Choose $A^i_t$ using $O^i_t$ and the policy derived from $Q^i$}
                        \State Take action $A^i_t$ in the environment
                        \State Receive and store $R_{t+1}^{}$ and $O^i_{t+1}$, with $O^i_{t+1} \in S_{t+1}$, from environment
                        \State $\sum_{f=0}^{P-1} R_{t+1+f}$ \Comment{Calculate the FAR based on the rewards for the next round of actions resulting from $A^i_t$}
                        \State $Q^i(O^i_t, A^i_t) \gets Q^i(O^i_t, A^i_t) + \alpha \bigr[ \sum_{f=0}^{P-1} R_{t+1+f} + \gamma \max_{a} Q^i(O^i_{t+P}, a) - Q^i(O^i_t, A^i_t) \bigr]$
                        \State $t \gets t + 1$
                    \EndFor
                \EndWhile
            \EndFor
        \end{algorithmic}
    \end{algorithm}  

    Due to the turn-based nature of the environment, for each episode, the algorithm steps through each agent (line 6) until the global terminal state $S_{t+1}$ is reached. Since
    the algorithm depicts Q-learning, the agents use an $argmax$ function to select an action based on the Q-values for a given observation, shown in line 7. After an action has
    been chosen, it is sent to the environment, which reacts accordingly and produces a new observation and reward. Note that due to the unique nature of turn-based settings, a
    special type of forward accumulated reward (FAR) must be used which consists of the one round return resulting from each agent's action~\cite{mappo}, i.e. $\sum_{f=0}^{P-1}
    R_{t+1+f}$. Finally, the original observation, next observation, as well as FAR are used in line 11 to update the action-value function according to equation
    (\ref{update_step_ql}).

    Instead of distinct policies $Q^i$, independent Q-learning agents can make use of a shared policy $Q$, especially when the environment contains symmetries~\cite{iql_og}. This
    allows the update step in equation (\ref{update_step_ql}) to be updated more frequently by using the experiences of all the independent agents. This has proven to offer
    significant performance gain and allow agents to learn more effectively~\cite{iql_og}.

\subsection{Deep Q-learning}\label{DQN} 
    Deep Q-learning is an extension of tabular Q-learning where artificial neural networks (referred to as deep Q-networks or DQNs) are used as non-linear function
    approximators~\cite{dqn_main}. Deep Q-learning implements an experience replay memory to store the experiences $e = \langle S_t, A_t, R_{t+1}, S_{t+1}
    \rangle$~\cite{replay_mem}, which in turn is sampled in random batches $b$ to remove correlation within the observation sequence, and thereby smoothing over the changes within
    the data distribution~\cite{dqn_main}. 
    
    Deep Q-learning usually incorporates a \textit{policy network} and a \textit{target network}~\cite{dqn_main}. The policy network is used to select actions and utilises the
    update step, while the target network serves as a baseline when calculating the \revision{\textit{TD-error}}. The target network is updated periodically with the policy network
    to reduce correlation with the target. The weights of the policy network are updated using backpropagation with the goal of minimizing the \revision{TD-error}, which is defined
    as
    \begin{equation}\label{td error}
        \begin{aligned}
            L_j(\theta_j^{}) = {}  & \mathbb{E}_{b}[(R_{t+1} + \gamma \max_{a}Q(S_{t+1}, a;\theta_j^-) - Q(S_t, A_t; \theta_j^{}))^2],
        \end{aligned}
    \end{equation}
    where $\theta_j^{}$ and $\theta_j^-$ are the weights of the policy-and target network, respectively, at iteration $j$~\cite{dqn_main}. In practice the backpropagation and
    calculation of the \revision{TD-error} is usually handled by an optimizer, such as the \textit{Adam} optimizer~\cite{adam_opt}. Similar to tabular Q-learning, deep Q-learning
    can be extended to MARL using independent deep Q-learning~\cite{idqn_pong}, and can also make use of a shared policy. 
    
\subsection{Options}\label{options}
    Options are built on the concept of semi-MDPs, where actions can have varying lengths over multiple time steps $t$ as the environment transitions from $S$ to
    $S'$~\cite{sutton1999options}. When an agent choses an option $\omega$, the option executes stochastically according to a policy $\pi_\omega$. This policy allows for existing domain
    knowledge to be incorporated into RL techniques. For example, in the case of a robot navigating terrain, a potential option can include: \{move around boulder\}, and once the
    agent choses that option, an object avoidance controller takes over executing policy $\pi_\omega$. 
    
    Once policy $\pi_\omega$ has concluded, the agent can choose a new option, effectively simplifying the problem at hand by solving it on a higher level, since the agent no
    longer has to focus on learning the \emph{primitive actions}\footnote{Primitive actions refer to the raw actions of the environment, for example: move up, move left, move right,
    move down, etc.} as well as the overarching problem simultaneously. Thus, an option $\omega$ is defined as
    \begin{equation}\label{option}
        \langle I_\omega, \pi_\omega, \beta_\omega \rangle,
    \end{equation}
    where $I_\omega$ is the initial condition that must be met for an option to become available in a certain state ($S_t \in I_\omega$), $\pi_\omega$ the policy according to which
    the option executes, and $\beta_\omega$ the termination condition to which the option is executed~\cite{sutton1999options}. The policy $\pi_\omega$ can also be defined using
    RL, a process known as intra-option learning~\cite{sutton1999options}. This has further lead to the development of \emph{option discovery}, which uses a Laplacian framework to
    allow an agent to define and learn its own options as well as the intra-option policies while interacting with the environment~\cite{option_discovery}. Options have also been
    extended to MARL with Amato \textit{et al.}~\cite{macdec_pomdp} introducing their macro-action Dec-POMDP (MacDec-POMDP) and showing how options affect the Dec-POMDP structure. 
    
    Recently, Chen \textit{et al.}~\cite{option_discovery_marl} explored option discovery within MARL and introduced their own variation on the Laplacian framework using Kronecker
    graphs. Options within MARL poses unique challenges due to the temporal disconnect of the actions as a direct result of them having varying lengths. The actions are also chosen
    asynchronously, which further adds to the complexity and non-stationarity of the environment~\cite{macdec_pomdp}. Additionally, agents in a synchronous setting must often wait
    for other agents to complete their options before they can initiate a new option, especially if that new option is a cooperative one~\cite{macdec_pomdp}. 
    
\section{Augmenting the Action Space with \emph{Conventions}}\label{conventions}
    Similar to how options enabled the incorporation of existing domain knowledge into RL algorithms by modifying an agent's action space, we propose a MARL solution strategy to
    achieve a similar goal. Options required the defining of semi-MDPs, where actions can have varying lengths, and in a MARL scenario will theoretically require decentralised
    partially observable semi-MDPs (encapsulated by Amato \textit{et al.'s} MacDec-POMDP~\cite{macdec_pomdp}). However, our solution strategy does not directly change the length of
    an action by having an external policy take over, rather it facilitates implicit communication through a specific behaviour as a result of a sequence of actions, based on
    predefined principles, called a \emph{convention}. Each \emph{convention} contains \emph{convention}-steps that span over and include multiple time steps and multiple agents,
    requiring \revision{each agent} to actively opt in for the \emph{convention} to reach fruition. Therefore, our augmentation of the action space does not directly change the
    Dec-POMDP framework, and merely acts as an extension of the existing action space to allow for more advanced behaviours.
    
    Ideally an agent's action space can be comprised of a pure \emph{convention} space, i.e. containing only \emph{convention}-steps and no primitive actions. However, there can
    exist scenarios where no \emph{convention}-steps are available, and the agents will not be able to take any actions. To solve this problem, we propose augmenting the action
    space with primitive actions \textbf{and} \emph{conventions} simultaneously, similar to the discussion by Sutton \textit{et al.}~\cite{sutton1999options} on combining primitive
    actions and options at the slight cost of performance. We will prove in Section~\ref{small_hanabi_results} at the hands of the Small Hanabi problem, that this solution strategy
    only slightly impacts the overall performance when comparing agents with a pure \emph{convention} space and an augmented action-\emph{convention} space. 

\subsection{\emph{Conventions}}
    Existing MARL approaches in partially observable environments, such as Hanabi, naturally develop conventions, but these conventions are nonsensical and vary greatly from run to
    run~\cite{otherplay,sp_vs_cp_1,sp_vs_cp_2}. We propose incorporating existing human conventions into MARL algorithms using artificial \emph{conventions}, which act as advanced
    cooperative actions that span over and include multiple time steps and agents. \emph{Conventions} require that all the agents involved participate or ``subscribe'' in order for
    it to reach fruition, however the agents cannot communicate directly which \emph{convention} is being started or followed.

    \emph{Conventions} can be divided into two categories, namely available \emph{conventions} and active \emph{conventions}, based on if an agent can initiate the
    \textit{convention} or continue the \textit{convention}. Each \emph{convention} contains a number of steps for that \emph{convention} (referred to as \emph{convention}-steps)
    which corresponds to the number of actions in that \emph{convention}, allowing it to reach fruition and result in a desired behaviour. We define a \emph{convention} $c_k$,
    containing a number of \emph{convention}-steps $m_k$, as
    \begin{equation}\label{conventions_eq}
        c_k = \langle m_k, \lambda_k^1, \pi_k^1, \lambda_k^2, \pi_k^2, \dots, \lambda_k^{m_k}, \pi_k^{m_k} \rangle.
    \end{equation}
    The set of conditions $\lambda_k$ determine which \emph{convention}-steps are currently available to an agent based on their current observation, i.e. where $O_t \in \lambda_k$
    determines the available \emph{convention}-steps $1:m_k$. Subsequently, the set of policies $\pi_k$ determine the corresponding environment action $A_t$ for each
    \emph{convention}-step with $A_t = \pi_k(O_t)$. 
    
    Thus, the initial condition $\lambda_k^1$ must be met for a \emph{convention} to become available, and if an agent chooses to initiate $c_k$, the policy $\pi_k^1$ determines
    the corresponding environment action to start the \emph{convention}. The \emph{convention} will then become active, and agents can ``subscribe'' to $c_k$ if they meet any of
    the conditions $\lambda_k^{2:m_k}$, with $\pi_k^{2:m_k}$ determining their environment action to continue $c_k$. Eventually an agent will have the opportunity to perform a
    unique continuation action to complete $c_k$, but only if they meet the final condition $\lambda_k^{m_k}$, with $\pi_k^{m_k}$ determining their environment action to terminate
    $c_k$ and allow it to reach fruition. When implementing $K$ \emph{conventions} in an agent's action space, the \emph{convention}-steps of the different \emph{conventions} are
    combined to form the \emph{convention}-step space\footnote{Note that the number of implemented \emph{conventions} $K$ is not equal to the size of the \emph{convention}-step
    space, since \emph{conventions} are variable length vectors based on the number of steps $m_k$ contained within each \emph{convention} $c_k$.} defined as
    \begin{equation}
        C = \langle \pi_0^1, \pi_0^2, \dots, \pi_0^{m_1}, \pi_1^0, \pi_1^1, \dots, \pi_K^{m_K} \rangle,
    \end{equation}
    with
    \begin{equation} \label{conv_size}
        |C| = \sum_{j=0}^K m_j.
    \end{equation}
    
    \emph{Conventions} are further distinguished based on their step-size $m_k$, with single-step \emph{conventions} having a step-size of $m_k=1$, two-step \emph{conventions}
    having $m_k=2$, and multistep \emph{conventions} having $m_k > 2$. Single-step \emph{conventions}, i.e. $c_k = \langle 1, \lambda_k^1, \pi_k^1 \rangle$, contain only one
    condition $\lambda_k^1$, as well as one policy $\pi_k^1$, which simultaneously acts as the initial-and final conditions of the \emph{convention}, allowing an agent to initiate
    and terminate the \emph{convention} on a single time step. Although this might seem similar to a primitive action, single-step \emph{conventions} still allow for additional
    communication through the observed behaviour resulting from the policy $\pi_k^1$, since the policy will result in a different environment action based on the current
    observation, as opposed to primitive actions which remain constant independent of the current observation. Two-step \emph{conventions} only contain two conditions and two
    policies, which act as the initial-and terminating steps of the \emph{convention}, i.e. $c_k = \langle 2, \lambda_k^1, \pi_k^1, \lambda_k^2, \pi_k^2 \rangle$. 

    It is important to note that \emph{conventions} do not have to be sequential, e.g. if agent 3 had their turn in between agent 1 and 2, and \textit{convention} $c_1$ is active
    and only relevant to player 1 and 2, agent 3 would be able to initiate or subscribe to a different \emph{convention} and $c_1$ will remain active. When more \emph{conventions}
    are introduced, the learning problem becomes apparent and crucial, since agents must learn which \emph{convention} is optimal in certain scenarios where more than one
    \emph{convention} is available. Furthermore, since agents have the ability to opt in to a \emph{convention}, they can also choose not to, which will result in an active
    \emph{convention} terminating without a completing action. This allows agents to halt certain \emph{conventions} based on their unique observations alluding to it being
    non-optimal. Thus, the agents must learn when to prioritise superior \emph{conventions} over existing or active inferior ones. We note that the majority of our implemented
    \emph{conventions}, presented and discussed in Appendix~\ref{imp_convs}, are single-and two-step \emph{conventions}, with the only multistep \emph{conventions} being the
    \textit{Prompt} and the \textit{Finesse} (each having an $m_k = 3$).  

\subsection{Action Space Augmentation}\label{aug}
    \emph{Conventions} can either act as the only actions available to an agent, or can be incorporated into the existing action space using action space augmentation. This
    requires that the \emph{convention}-step space be appended to the existing primitive-action space, and produce the augmented action-\emph{convention} space. However, before
    this can be achieved, the primitive actions must be translated into unique \emph{conventions} to ensure effective cohesion. Ultimately, a primitive action is a special type of
    single-step \emph{convention} with no initial condition and a fixed deterministic policy mapping to a certain environment action independent of the current observation, i.e. 
    \begin{equation}\label{prim_act_conv}
        c_k' = \langle 1, \lambda_k^1 = \{\mathcal{O}\}, \pi_k^1 = a_k \rangle,
    \end{equation}
    where $\mathcal{O}$ represents any observation, and $a$ the corresponding primitive action.
    
    Thus, we can combine the primitive actions $c_k'$ defined in equation (\ref{prim_act_conv}) and the \emph{conventions} $c_k$ defined in equation (\ref{conventions_eq}) to
    produce the augmented action-\emph{convention} space $C$ defined as
    \begin{equation}\label{action_conventions_eq}
        \begin{aligned}
            C {}    & = \{c_0', c_1', \dots, c_{|A|-1}', c_{|A|}', c_{|A|+1}, \dots, c_{|A|+K}\} \\
                    & = \langle \pi_0^1, \pi_1^1, \dots, \pi_{|A|-1}^{1}, \pi_{|A|}^1, \pi_{|A|+1}^1, \pi_{|A|+1}^2, \dots, \pi_{|A|+K}^{m_K-1}, \pi_{|A|+K}^{m_K} \rangle,
        \end{aligned}
    \end{equation}
    with
    \begin{equation}
        |C| = \sum_{j=0}^K m_j + |A|.
    \end{equation}
    This ensures that an agent has access to actions in situations where \emph{conventions} aren't applicable or available. Additionally, this increases the learning
    problem complexity since an agent must learn to effectively utilise \emph{conventions} as well as primitive actions, given scenarios where both are applicable and available. 
    
    To implement action space augmentation with \emph{conventions}, a few changes must be made to the existing algorithm of interest. In the case of independent Q-learning shown in
    Algorithm~\ref{ql algorithm}, the algorithm will receive an additional input detailing the list of implemented \emph{conventions}, which in turn gets appended to the existing
    primitive action space to produce the augmented action-\emph{convention} space defined in equation (\ref{action_conventions_eq}). The output remains the same, i.e. a policy in
    the form of an action-value function, however this function will be conditioned on the augmented action-\emph{convention} space $C$, rather than just the primitive action
    space $A$. Therefore, we can define Algorithm~\ref{ql aug act conv algorithm}, which acts as the algorithm for independent Q-learning with an augmented action-\emph{convention}
    space in a turn based environment.

    \begin{algorithm}[!h]
        \caption{Independent Q-learning with an augmented action-\emph{convention} space in a turn-based environment}\label{ql aug act conv algorithm}
        \begin{algorithmic}[1]
            \State Initialise hyperparameters similar to Algorithm~\ref{ql algorithm}
            \State Define the list of \emph{conventions} using $c_k = \langle m_k, \lambda_k^1, \pi_k^1, \lambda_k^2, \pi_k^2, \dots, \lambda_k^{m_k}, \pi_k^{m_k} \rangle$, with $k
            = 0:K$
            \State Translate the primitive actions to \emph{conventions} using $c_k' = \langle 1, \{\mathcal{O}\}, \pi_k^1 = a_k \rangle$, with $k = 0:|A|$
            \State Combine the primitive actions and \emph{conventions} to produce the augmented action-\emph{convention} space $C = \{c_0', c_1', \dots, c_{|A|-1}', c_{|A|}', c_{|A|+1}, \dots, c_{|A|+K}\}$
            \State Initialise all $Q(O, C) \gets 0$

            \For{each episode}
                \State Reset environment and set total time $t \gets 0$

                \While{$S_{t+1}$ is not terminal}
                    \For{all players $i$ from $0$ to $P-1$}
                        \State Obtain $O^i_t$ from environment 

                        \State ${C}_{mask} = \{C|O_t^i \in \lambda_k\}$ \Comment{Define the action mask using the available and active \emph{conventions} determined by the set of
                        conditions $\lambda_{k}$ and the current observation $O_t^i$} 
                        \State ${C}^i_t = \text{argmax}_{c} (Q^i(O^i_t, c) \cap {C}_{mask})$ \Comment{Choose ${C}^i_t$ from the available and active \emph{conventions} based on
                        $O^i_t$, the action mask ${C}_{mask}$, and the policy derived from $Q^i$}

                        \State $m, k = C[{C}^i_t]$ \Comment{Index the augmented action-\emph{convention} space to determine $m$ and $k$}
                        \State $A^i_t = \pi_k^{m}(O^i_t)$ \Comment{Use the \emph{convention} policy to determine the environment action}

                        \State Take action $A^i_t$ in the environment
                        \State Receive and store $R_{t+1}^{}$ and $O^i_{t+1}$, with $O^i_{t+1} \in S_{t+1}$, from environment
                        \State $\sum_{f=0}^{P-1} R_{t+1+f}$ \Comment{Calculate the FAR similar to Algorithm~\ref{ql algorithm}}
                        \State $Q^i(O^i_t, {C}^i_t) \gets Q^i(O^i_t, {C}^i_t) + \alpha \bigr[ \sum_{f=0}^{P-1} R_{t+1+f} + \gamma \max_{c} Q^i(O^i_{t+P}, c) - Q^i(O^i_t, {C}^i_t) \bigr]$
                        \State $t \gets t + 1$
                    \EndFor
                \EndWhile
            \EndFor
        \end{algorithmic}
    \end{algorithm}  
    
    In line 2, we define the list of \emph{conventions} from $0:K$, followed by the translation of the primitive actions to the special form of single-step \emph{conventions}
    $c_k'$ in line 3. The \emph{conventions} and primitive actions are then be combined to form the augmented action-\emph{convention} space $C$, shown in line 4. In line 11, the
    current observation, produced by the environment for the active player $i$ at time $t$, determines the available and active \emph{conventions} using the set of conditions
    $\lambda_{k}$ for each \emph{convention}. This produces the action mask to indicate which \emph{conventions}-steps are currently available to the agent, and is often used in
    deep learning to ensure that an agent choses ``legal'' actions given a certain state. Notably, this also encapsulates the legal primitive actions given the current observation,
    since they are contained within the augmented action-\emph{convention} space. Once the action mask is determined, agent $i$ can start or continue an available or active
    \emph{convention} (or take a primitive action) by choosing a \emph{convention}-step $C_t^i$ based on the policy derived from $Q^i$, the action mask ${C}_{mask}$, and their
    observation $O^i_t$, shown in line 12.
    
    In line 13, we use the chosen \emph{convention}-step $C_t^i$ to determine the corresponding \emph{convention} $k$, i.e. the \emph{convention} which contains the specific
    \emph{convention}-step, and its appropriate step $m$, i.e. the step in that \emph{convention} where the chosen \emph{convention}-step occurs. The \emph{convention} $k$ and the
    step $m$ can then be used to determine the appropriate environment action $A_t^i$, according to the policy $\pi_k^m$ and the current observation $O_t^i$, shown in line 14. If
    the agent chose a primitive action from the augmented action-\emph{conventions}, the step $m$ will be equal to one and the \emph{convention} $k$ will correspond to the
    primitive action $a_k$, according to equation (\ref{prim_act_conv}). Finally, the environment action $A_t^i$ is sent to the environment which in turn produces the next
    observation and reward used to calculate the FAR, similar to Algorithm~\ref{ql algorithm}. As is the case with human players, the agents aren't allowed to communicate which
    \emph{convention} is currently being started or followed, but rather only observes the environment action determined by the appropriate \emph{convention} policy, shown in line
    15.

\begin{figure}[!h]
        \centerline{\includegraphics[width=\textwidth]{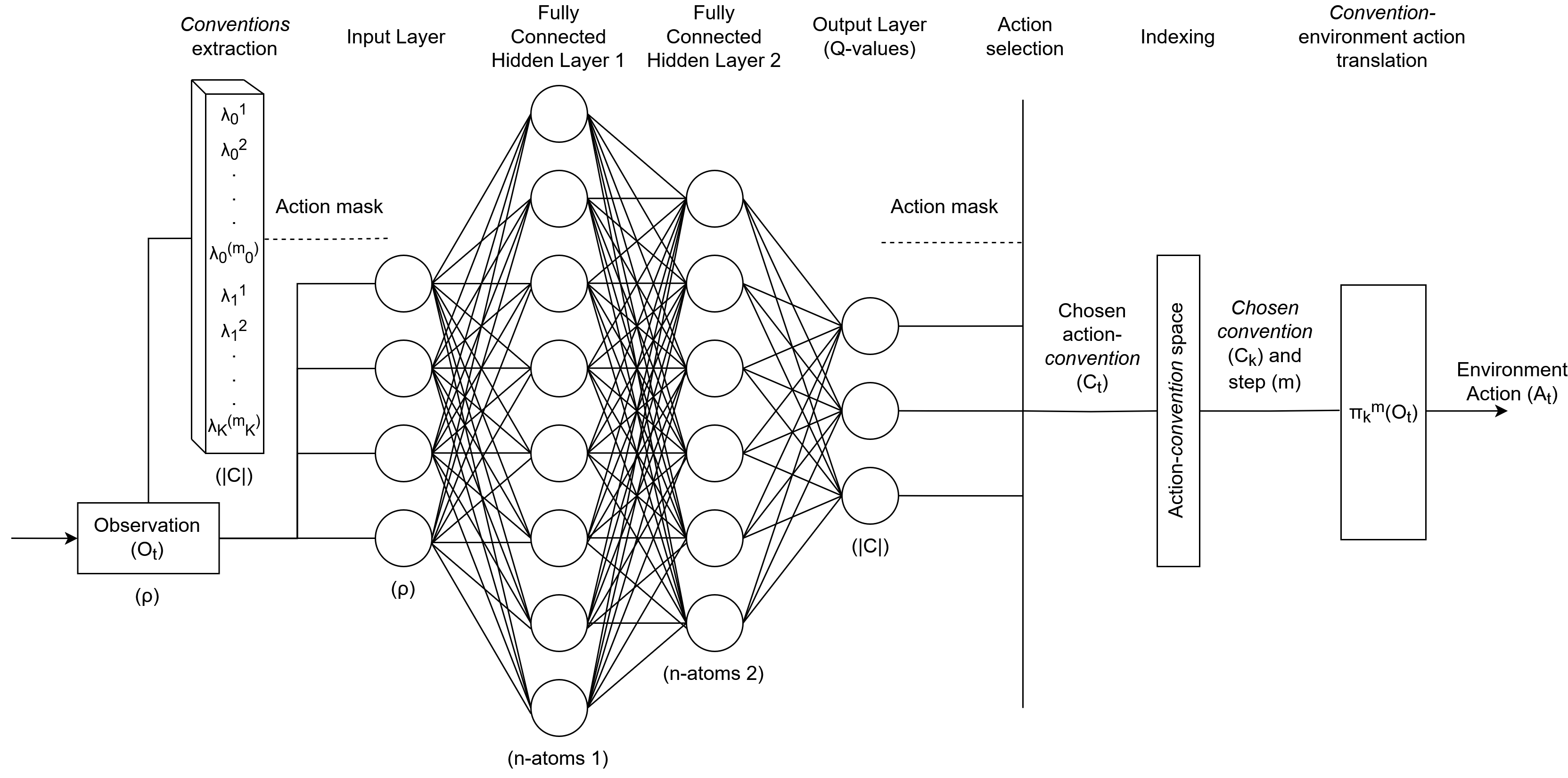}} \caption{Architecture design for a feed forward neural network (NN) with an augmented
        action-\emph{convention} space applied. The environment provides the observation $O_t$ and receives the environment action $A_t$. The size of the input layer is equal to
        the size of the observation tuple ($\rho$), and each hidden layer has a size equal to the number of atoms for that layer. The size of the output layer is equal to the size
        of the augmented action-\emph{convention} space ($|C|$), and the action selection is determined by the MARL algorithm. Finally, the chosen augmented
        action-\emph{convention} determines the specific \emph{convention} $c_k$ and its step $m$, which in turn is used to produce the environment action $A_t$ according to the
        policy $\pi_k^m$ and the observation $O_t$.}\label{arch}
    \end{figure}

    From the environment's perspective, agents are merely choosing actions based on the observations the environment provided, but in reality the agents are choosing these actions
    based on a set of predefined \emph{convention} policies $\pi_k$. In the case of deep RL, lines 11 and 13-14 indicate that our approach requires two new layers to be added to
    the agent's architecture, namely one before the input layer to determine the available and active \emph{conventions}, i.e. the action mask, and one at the output to translate
    the chosen \emph{convention}-step $C_t^i$ into an environment action, based on the corresponding \emph{convention} $k$, the \emph{convention's} step $m$, the corresponding
    policy $\pi_k^m$, and the current observation $O_t$. This is illustrated in Fig.~\ref{arch}, and in the case of deep Q-learning, the action selection step is performed using
    $\text{argmax}_{c} Q(O_t, c)$, with the available and active \emph{conventions} as well as the primitive actions forming the action mask.

    Agents with an augmented action-\emph{convention} space can also make use of a shared policy, i.e. each agent will have access to the same network architecture depicted in
    Fig.~\ref{arch}, as discussed in Section~\ref{iql}. Since \emph{conventions} act as an extension of the existing action space, it can be applied to any MARL method by adding
    the necessary \emph{convention} components on top of the existing architecture as well as translating the existing primitive action space into single-step \emph{conventions},
    similar to our discussion on independent Q-learning. 

\section{Hanabi}\label{hanabi} 
    To test the capabilities of \emph{conventions}, we will use the cooperative Hanabi environment, which was proposed as a frontier for MARL algorithm development by Bard
    \textit{et al.}~\cite{hanabi_ai}. Hanabi is a 2--5 player card game, best described as a cooperative solitaire~\cite{hanabi_ai}. Players have a hand of five cards for
    \revision{two-and three-player}, and four cards for \revision{four-and five-player}. Each card has a suit/colour (red, yellow, green, white, or blue) and a rank (1 to 5). The
    deck comprises 50 cards total, with 10 cards for each suit with a distribution of three 1's, two 2's, 3's, and 4's, and finally only one 5. The aim of the game is to stack
    these suits in numerical order from 1 to 5, however players cannot see their own cards (i.e. their hands are hidden), but can see the cards of their fellow players.
    
\begin{figure}[h]
        \centerline{\includegraphics[width=28pc]{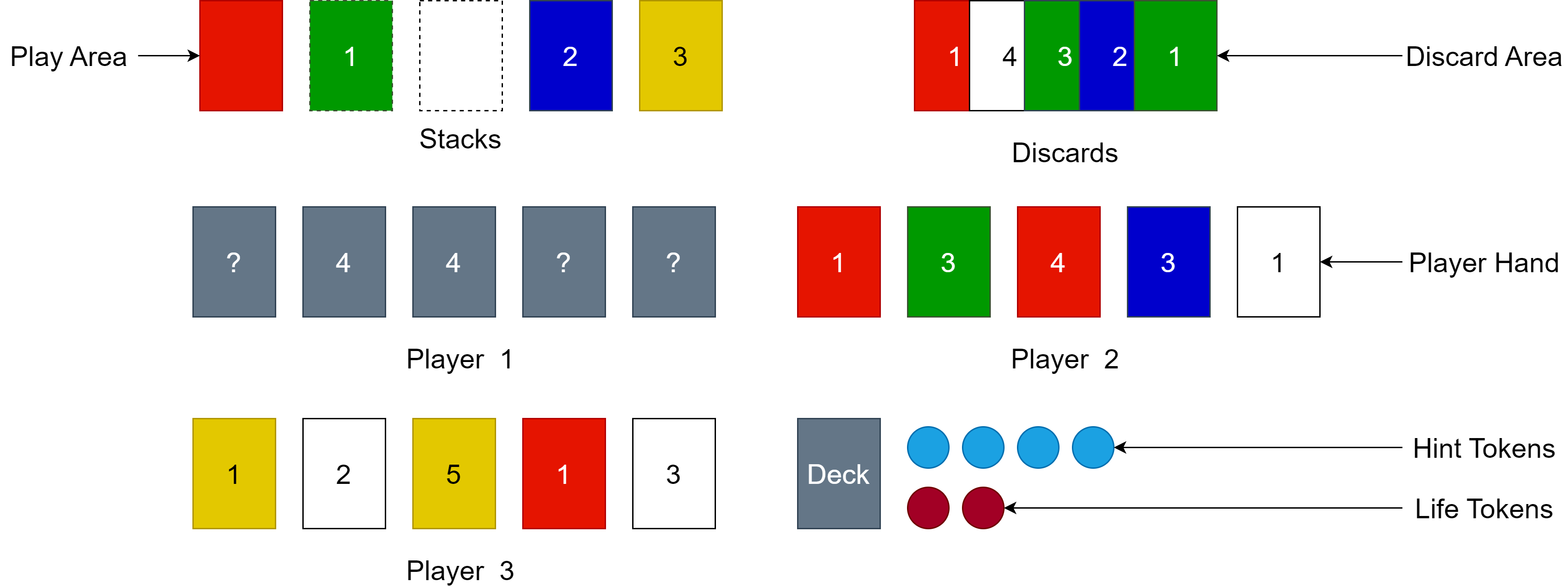}} \caption{An example game of Hanabi as seen from the perspective of player 1. There are four hint tokens
        left\revision{,} and the players have lost one of their shared life tokens. Player 1 knows about two 4s in their hand and the green, blue and yellow stacks have been
        partially completed, leading to a current game score of 6/25. It is now player 1's turn, and they can take the hint (to player 2 or 3), play (from their hand) or discard
        (from their hand) action.}\label{han}
    \end{figure}

    The players take consecutive turns, and on each turn a player can either \emph{play}, \emph{discard}, or \emph{hint}. \emph{Hinting} involves revealing all the cards in another
    player's hand matching a certain rank \textit{or} a certain suit, and consumes one of the limited (and shared) hint tokens. In the centre is a stack for each suit where the
    players must \emph{play} their cards, and a successful play entails playing a card that follows the current card on top of a stack (starting at 0). If the card played does not
    follow the current card on a stack, the play was unsuccessful (called a \emph{misplay}) and the players lose one of their shared life tokens. \emph{Discarding} involves
    removing a card from the current player's hand and adding it to the discard pile, effectively removing it from the game, while also replenishing a hint token. An example game
    is shown in Fig.~\ref{han}.

    At the start of the game the players have a shared total of three life tokens and eight hint tokens. If all the hint tokens are depleted, a player cannot take the hint action
    and must either play or discard a card from their hand. The players are awarded a shared score depending on the number of cards successfully placed on each stack. If the
    players manage to build the stacks to the maximum of 5 each, the game ends and the players receive a perfect score of 25/25. Alternatively, if the players lose all of their
    life tokens (called ``bombing out''), or if the deck has been depleted, the game also ends. Note, bombing out results in a score of 0/25 independent of how many cards were
    played successfully, while deck depletion results in a final score equal to the current score. A discard action, or a misplay, will result in the card being removed from the
    game, and that player must draw a new card from the deck to replenish the missing card (which is also hidden from that player). 

    Due to the partial observability imposed by a player's hand being hidden, as well as the limited communication channel in the form of hinting, Hanabi offers a very interesting
    and useful challenge for MARL agents to solve~\cite{hanabi_ai}. Actions are highly correlated and players must coordinate effectively to achieve success, often requiring
    advanced strategies and reasoning over the intentions of other player's actions to reliably beat the game.  

\subsection{Human Conventions in Hanabi}
    Due to the nature of Hanabi, players have too few hint tokens to effectively convey all the needed information (colour and rank) of the 25 playable cards~\cite{hanabi_ai}. This
    becomes even more challenging as the number of players increase and hint tokens become more valuable, as well as the number of playable turns that keep decreasing\footnote{As
    the number of players increase, the cards remaining in the deck decrease, and a single player has less playable turns.}. Hints must therefore be used wisely, and must convey
    information about more than just one card. This is possible since more than one card with a similar rank or suit can be touched (or revealed) by a single hint, along with
    negative information\footnote{Negative information refers to implied information about non-focus cards, e.g. if a player hints that two cards are green, the other cards must
    therefore be non-green.} being given about all the other cards in a hand. Unfortunately, this is not enough, and players who don't use additional strategies will struggle to
    reliably achieve a score above 15/25.

    Players, therefore, require a way to convey additional information through implicit communication, while still remaining within the rules of the game. As stated by Bard
    \textit{et al.}~\cite{hanabi_ai}: ``This implicit information is not conveyed through the impact that an action has on the environment (i.e., what happens) but through the very
    fact that another player decided to take this action (i.e., why it happened)''. Players can then reason over the actions of others, and implicitly communicate through them.
    This is done through the use of conventions, which are built on principles or mutually agreed upon ``rules'', that players develop outside the game. For Example:
    
    \textbf{Conventions Example 1.} Let's assume the players in Fig.~\ref{han} have established a convention which states that if an ambiguous hint
    is given, the focus of that hint is on the left-most card. Therefore, using this convention, player 1 can hint to player 2 that they have two
    red cards in position one and three respectively (from the left), and player 2 will know (as a result of their convention) that the left-most card must be the
    playable red 1\footnote{This is a common human convention and is based on the ``Single card focus'' principle as discussed by the \href{https://hanabi.github.io/}{H-Group}
    and the implemented conventions found in Appendix~\ref{imp_convs}.}. 
    
    Since the game's release, players have developed their own intricate and extensive conventions, often specific to their group of friends or cooperators, examples include the
    \href{https://forum.boardgamearena.com/viewtopic.php?t=5252}{Board Game Arena}\footnote{https://forum.boardgamearena.com/viewtopic.php?t=5252} and the
    \href{https://hanabi.github.io/}{H-Group}\footnote{https://hanabi.github.io/}, with the \href{https://hanabi.github.io/}{H-Group} being the most widely used and adapted.

\subsection{Artificial Conventions in Hanabi}
    To incorporate existing human conventions into MARL, we use \emph{conventions} as described in Section~\ref{conventions}. This process is best described at the hand of an
    example:

    \textbf{Conventions Example 2.} In Fig.~\ref{han} each player is controlled by an agent, with the current player (referred to as the active player) being agent 1. The
    convention at hand (as mentioned in \textbf{Conventions Example 1}), which we will refer to as $c_1$, states that if an ambiguous hint is given, the focus of that hint is the
    left-most (or newest) card. Since the condition $\lambda_1^1$ is met, i.e. player 2 has two cards in their hand that share the same colour or rank and the left-most one is
    playable, the action \{start $c_1$\} is available for agent 1. Agent 1 chooses to initiate $c_1$ which in turn gets translated by the policy $\pi_{1}^1$ to an environment
    action of \{hint red to player 2\}, ending player 1's turn. It is now player 2's turn, and they have just received an ambiguous colour hint from player 1, and given their
    current observation they can derive that $c_1$ is currently active. This triggers the subscribing (and also completing) condition $\lambda_1^2$ (\revision{the condition} is
    completing since $m_1 = 2$), where agent 2 can now opt in to continue (and complete) $c_1$. When agent 2 chooses to subscribe to $c_1$, their action is translated by the policy
    $\pi_{1}^2$ to an environment action of \{play card in position 1\}, subsequently ending player 2's turn and completing $c_1$. 

    As seen by this example, and discussed in Algorithm~\ref{ql aug act conv algorithm}, the agents are not allowed to convey directly which convention was started or followed, and
    merely observe the behaviour of the other player through that player's environment action and their observation. We note that not all MARL scenarios allow for the observation
    of another agent's actions, and agents will therefore only make use of their observations when determining the \emph{conventions}. However, similar to the SMAC
    environment~\cite{smac}, Hanabi intrinsically allows the monitoring of other agents' actions, and is included in an agent's observation vector. Conventions Examples 1 and 2
    focused on the ``Single card focus'' principle, as discussed in Appendix~\ref{imp_convs_hanabi}, however there exists more advanced conventions that allow for improved implicit
    communication. These principles are all aimed at maximising the amount of information given by a single hint, in order to ensure there are enough hint tokens to effectively
    convey all the needed information of the 25 playable cards.

\subsection{Preliminary Results on Small Hanabi}\label{small_hanabi_results}
    To test the viability and capabilities of \emph{conventions}, we conduct initial tests using the Small Hanabi environment, since it offers a smaller state-action space and
    allows for a pure \emph{convention} space\footnote{A pure \emph{convention} space contains only \emph{conventions}, i.e. no primitive actions, and accounts for all possible
    scenarios, ensuring the agents will always have a \emph{convention} available. In Small Hanabi this is possible due to the small state-action space, but becomes a far greater
    challenge in the full Hanabi problem.}. We compare a pure \emph{convention} space and an augmented action-\emph{convention} space, as discussed in Section~\ref{aug}, against
    their primitive action benchmarks. Due to Hanabi having a long list of intricate conventions, in addition to augmenting the action space with \emph{conventions}, we experiment
    with the idea of simplifying the list of conventions during augmentation\footnote{This idea stemmed from the discussion by Sutton \textit{et al.}~\cite{sutton1999options},
    where they explored the capabilities of combining primitive actions with simplified options, since the primitive actions can account for the simplification at the slight cost
    of performance.}. This simplified list consists of the conventions that were the most straightforward and natural to develop, and were also the most common ones to occur in a
    game. The removed conventions can be considered edge-case or specific to certain game states that are very uncommon, and often \revision{result} in uncertainty regarding their
    effectiveness.

    Small Hanabi has a deck reduced to 20 cards with only two suits, is limited to only two players, and the player hand size \revision{is} reduced to two cards. Furthermore, the
    life tokens have been reduced to one, and the hint tokens to three, otherwise the game shares all the core mechanics as regular Hanabi. This results in the problem having a
    significantly smaller state-action space and allows for initial testing and development of algorithms, with the aim of solving the full Hanabi problem. Even though this problem
    is considered simpler than Hanabi, it is by no means an easy problem to solve, with a perfect score of 10/10 being difficult to achieve. Player hands are often ``locked'' with
    important cards that must not be discarded (called critical cards), hint tokens run out almost immediately, and players are often forced into no-win scenarios where they must
    choose a bad action or risk losing their single life token.

    In these tests, we use our in-house learning environment as well as the open-sourced learning environment developed by Bard \textit{et al.}~\cite{hanabi_ai}. We implement our
    own rudimentary conventions specific to Small Hanabi, which we developed through multiple human plays, presented and discussed in Appendix~\ref{imp_convs_small_hanabi}.
    Fig.~\ref{augmented_conventions_dqn} shows the result for independent Deep Q-learning (DQN) with a primitive-action space compared to independent Deep Q-learning with a pure
    \emph{convention} space. DQN with a pure \emph{convention} space performs significantly better than DQN with a primitive-action space, achieving a faster training time and
    improved converged performance. To validate these results, and improve on the performance further, we conduct an initial test using Rainbow and the open-sourced Hanabi learning
    environment with the Small Hanabi preset, the results are shown in Fig.~\ref{augmented_conventions_rainbow}. Rainbow is able to perform better than independent Deep Q-learning
    achieving a score of 7,6/10 compared to 5/10, and when substituting the primitive-action space for a pure \emph{convention} space, the agents achieve significantly faster
    training time along with a slight convergent performance increase. 

    \begin{figure}[!h]
        \centering
        \begin{subfigure}[b]{0.49\textwidth}
            \centering
            \includegraphics[width=\textwidth]{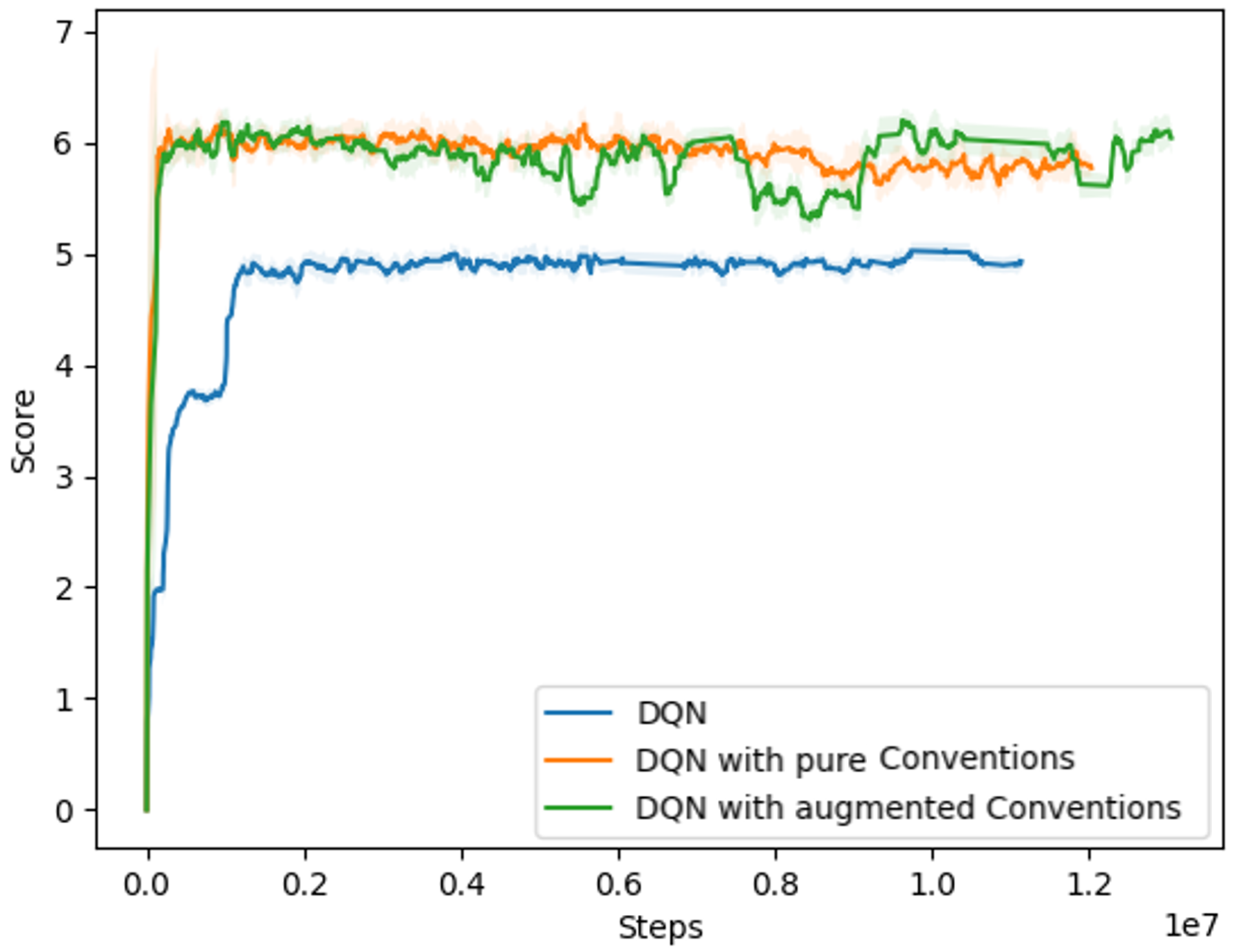}
            \caption{}\label{augmented_conventions_dqn}
        \end{subfigure}
        \hfill
        \begin{subfigure}[b]{0.49\textwidth}
            \centering
            \includegraphics[width=\textwidth]{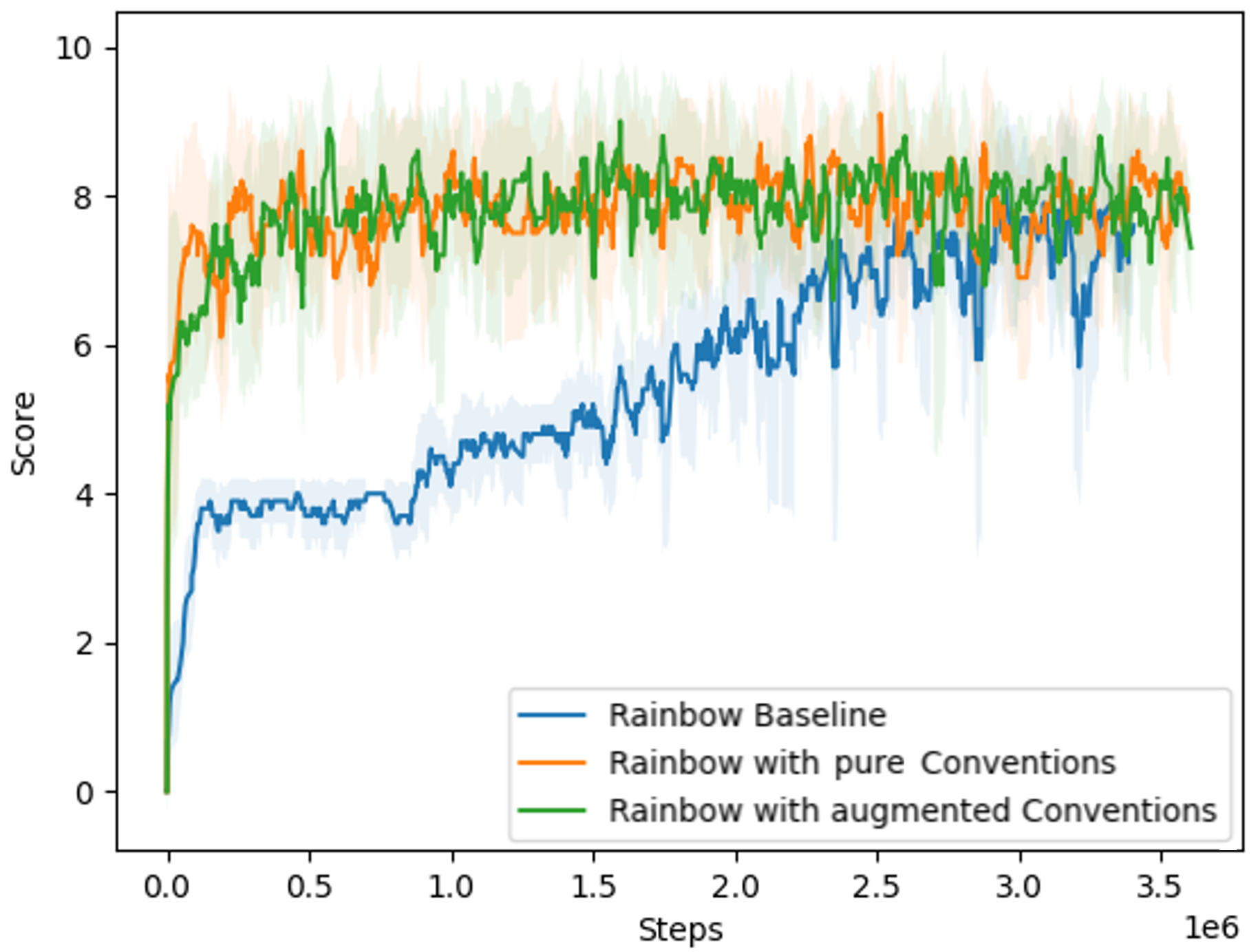}
            \caption{}\label{augmented_conventions_rainbow}
        \end{subfigure}
        \caption{(a) Learning curves for independent Deep Q-learning (DQN) with a primitive-action space, pure \emph{conventions} space, and an augmented and simplified
        action-\emph{convention} space tested in our in-house Small Hanabi environment. (b) Learning curves for Rainbow with a primitive-action space, pure \emph{conventions}
        space, and an augmented and simplified action-\emph{convention} space tested in \textit{DeepMind's} Hanabi learning environment with the Small Hanabi
        preset~\cite{hanabi_ai}. }\label{augmented_conventions}
    \end{figure}

    When augmenting the action space with \emph{conventions}, we removed \emph{conventions} 4-7 discussed in Table~\ref{small_hanabi_conv} found in
    Appendix~\ref{imp_convs_small_hanabi}. Fig.~\ref{augmented_conventions_dqn} and Fig.~\ref{augmented_conventions_rainbow} shows the performance for DQN agents and Rainbow
    agents, respectively, with an augmented action-\textit{conventions} space. The performance impact is almost insignificant when compared to each method's results for a pure
    \emph{convention} space. This demonstrates that even though only half of the \emph{conventions} are present alongside primitive actions, the agents are still able to perform
    significantly better compared to only having primitive actions, and only slightly impacts performance compared to pure \emph{conventions}. Therefore, when applying
    \emph{conventions} in the full Hanabi problem, only a subset of established human conventions will be needed alongside primitive actions. Most existing Hanabi conventions have
    a list of basic conventions which are built on the most fundamental principles, and are generally considered standardised within the community. It is these conventions, as
    presented and discussed in Appendix~\ref{imp_convs_hanabi}, alongside primitive actions that will be used when testing agents in the full Hanabi problem. 

    It is important to note that there could exist a combination of conventions that would result in better performance uplifts, especially when a larger variety of conventions are
    implemented. However for the sake of this argument, we focused on the basic conventions since they are generally the least contentious, less complex than more advanced
    strategies, and offer a good starting point for exploring the capabilities of action space augmentation with \emph{conventions} in Hanabi. Further research can be conducted to
    explore more advance \emph{conventions}, including advanced multistep \emph{conventions}, and their potential performance benefits.

\section{Performance Evaluation Using Hanabi}\label{results} 
    The preliminary tests for action augmentation with \emph{conventions} in Small Hanabi has shown promising results, and \revision{lead to the agents achieving} faster training
    \revision{times} and improved policies. We now shift focus to the full Hanabi problem for 2--5 players and with the conventions defined by the
    \href{https://hanabi.github.io/}{H-Group} and discussed in Appendix~\ref{imp_convs_hanabi}. Our results will focus on Bard \textit{et al.}'s~\cite{hanabi_ai} Rainbow agent as a
    baseline, and apply an augmented action-\emph{convention} space to improve on its performance. Rainbow was chosen as our baseline since it is among the methods that train the
    fastest, is the most sample efficient among the existing Hanabi agents~\cite{hanabi_ai,sad_hanabi,sad_search,mappo,otherplay}, has little run to run variance~\cite{hanabi_ai},
    is considered a widely applicable algorithm~\cite{rainbow_example_1,rainbow_example_2,rainbow_example_3}, and still leaves considerable performance to be desired within Hanabi.
    All tests are conducted on the open-sourced Hanabi learning environment developed by Bard \textit{et al.}~\cite{hanabi_ai}. 

\subsection{Experimental Setup}
    Before discussing the results, we will briefly highlight the architecture design for each deep RL technique. For a full list of the hyperparameters used in each method see
    Table~\ref{table of hypers 2} in Appendix \ref{hyperparams}. All methods receive a one-hot encoded observation defined by the environment as input to their neural networks,
    notably (by default) this observation tuple contains the most recent actions within the previous round, and is not added as an additional feature. The Rainbow agents use a
    Multilayer Perceptron (MLP) consisting of two feed forward neural networks with 512 neurons each, with the \textit{ReLU} activation function~\cite{relu} applied. It implements
    the Adam optimizer~\cite{adam_opt} to calculate the \revision{TD-error} and perform backpropagations to update the network weights. Additionally, the Rainbow agents use
    distributional reinforcement learning to predict the value distributions which are approximated as a discrete distribution over 51 uniform atoms, along with prioritise replay
    memory sampling~\cite{rainbow}. Even though Rainbow uses n-step bootstrapping, in these experiments a value of n=1 was found to be optimal, additionally noisy nets have been
    disabled in favour for a traditional decaying epsilon-greedy approach. 

    When applying \emph{conventions} to Rainbow, we only augment the action space and leave the reward signal and observation space untouched. Additionally, we add the necessary
    \emph{convention} layers to the network architecture, but keep the core algorithm unaltered, as illustrated by Fig.~\ref{arch}. Through testing, we found the optimal
    hyperparameters to be similar to baseline Rainbow, as shown in Table~\ref{table of hypers 2} in Appendix \revision{\ref{hyperparams}}. During evaluation, we compare each
    approach's exponential moving averages (with a weight value of 0.9995) and standard error of the mean. All agents are trained using an \textit{Intel} Core i7 10700K CPU and
    \textit{Nvidia} RTX 3080Ti GPU. The reward type for the Hanabi learning environment is set to non-lenient, i.e. the agents will receive a large negative reward (equal to the
    score) when bombing out. 

    Due to the symmetric nature of the environment, we apply a shared policy strategy during training, i.e. the agents share an action-value function that is updated based on a
    communal memory of each agent's experiences~\cite{central_decentral}. It is important to note that this still restricts learning by only using individual experiences, i.e.,
    there is no additional sharing of state information between each agent. During the cross-play evaluation of 2--5 player Hanabi, we chose samples from 10 separately trained
    agents for each player count, and show the results for the combination of agents that performed the best on average. 

\subsection{Self-play Performance}
\begin{figure}[!h]
        \centering
        \begin{subfigure}[b]{0.49\textwidth}
            \centering
            \includegraphics[width=\textwidth]{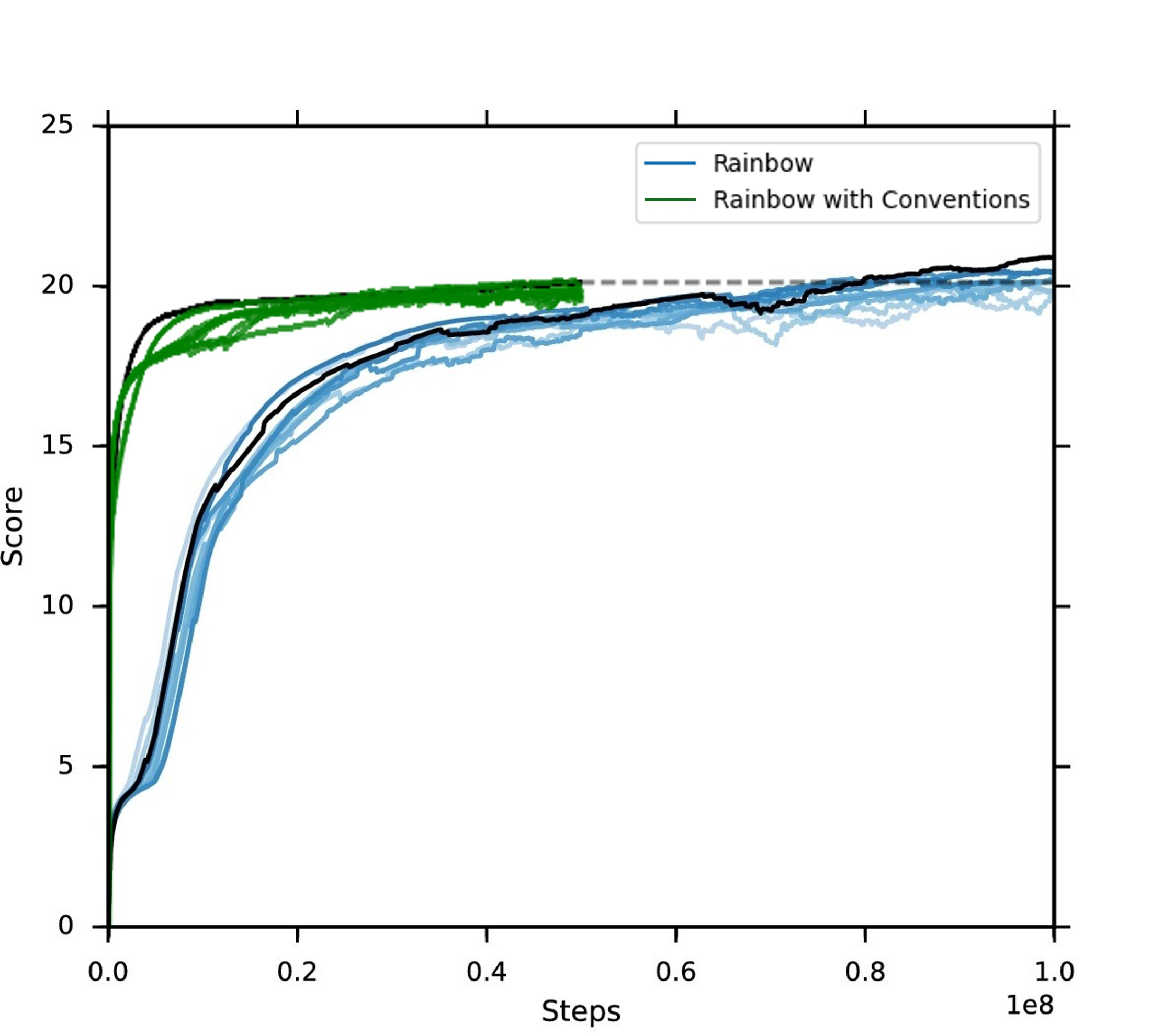}
            \caption{2-Player learning curves}\label{2p_lr}
        \end{subfigure}
        \hfill
        \centering
        \begin{subfigure}[b]{0.49\textwidth}
            \centering
            \includegraphics[width=\textwidth]{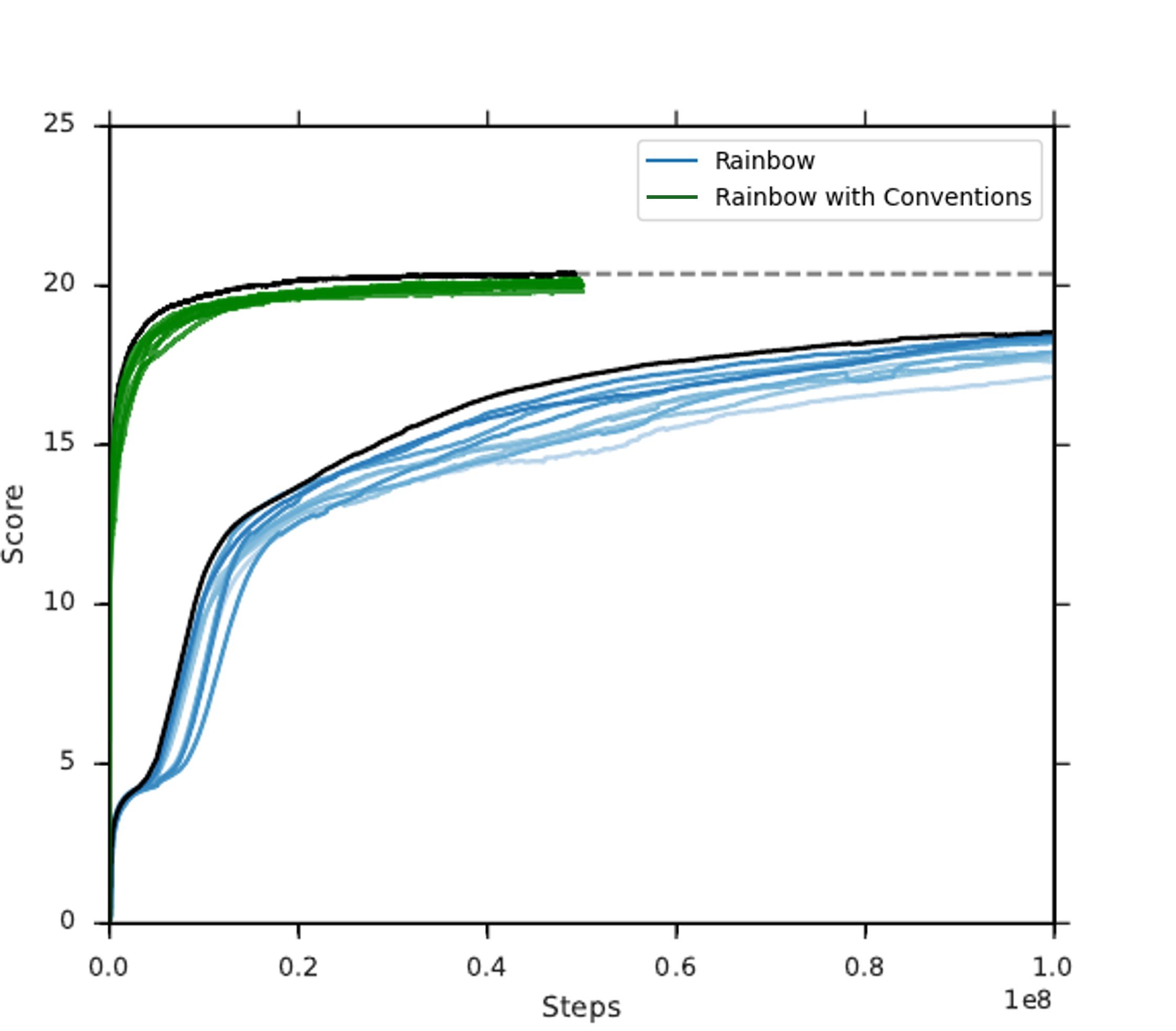}
            \caption{3-Player learning curves}\label{3p_lr}
        \end{subfigure}
        \hfill
        \centering
        \begin{subfigure}[b]{0.49\textwidth}
            \centering
            \includegraphics[width=\textwidth]{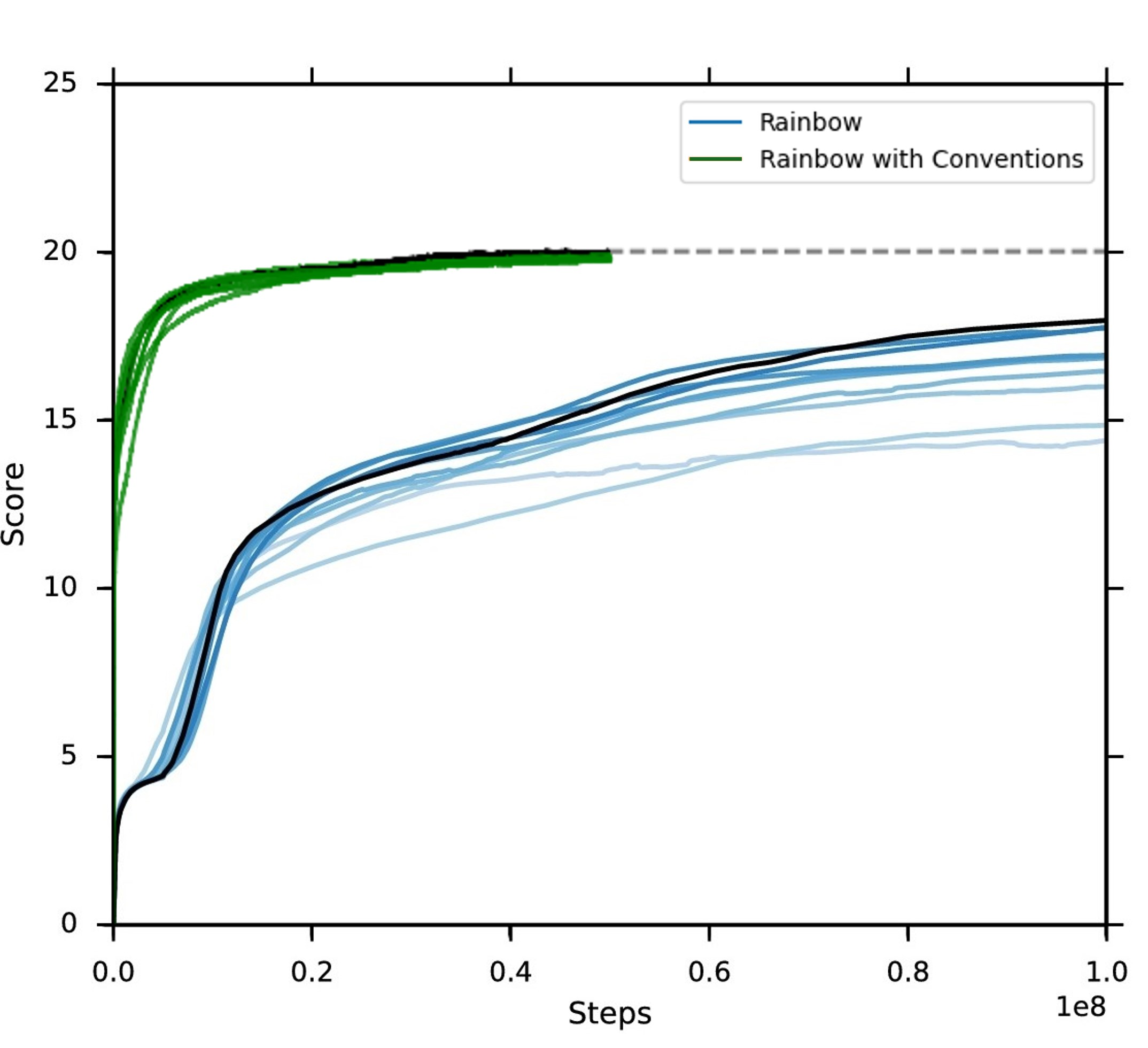}
            \caption{4-Player learning curves}\label{4p_lr}
        \end{subfigure}
        \hfill
        \centering
        \begin{subfigure}[b]{0.49\textwidth}
            \centering
            \includegraphics[width=\textwidth]{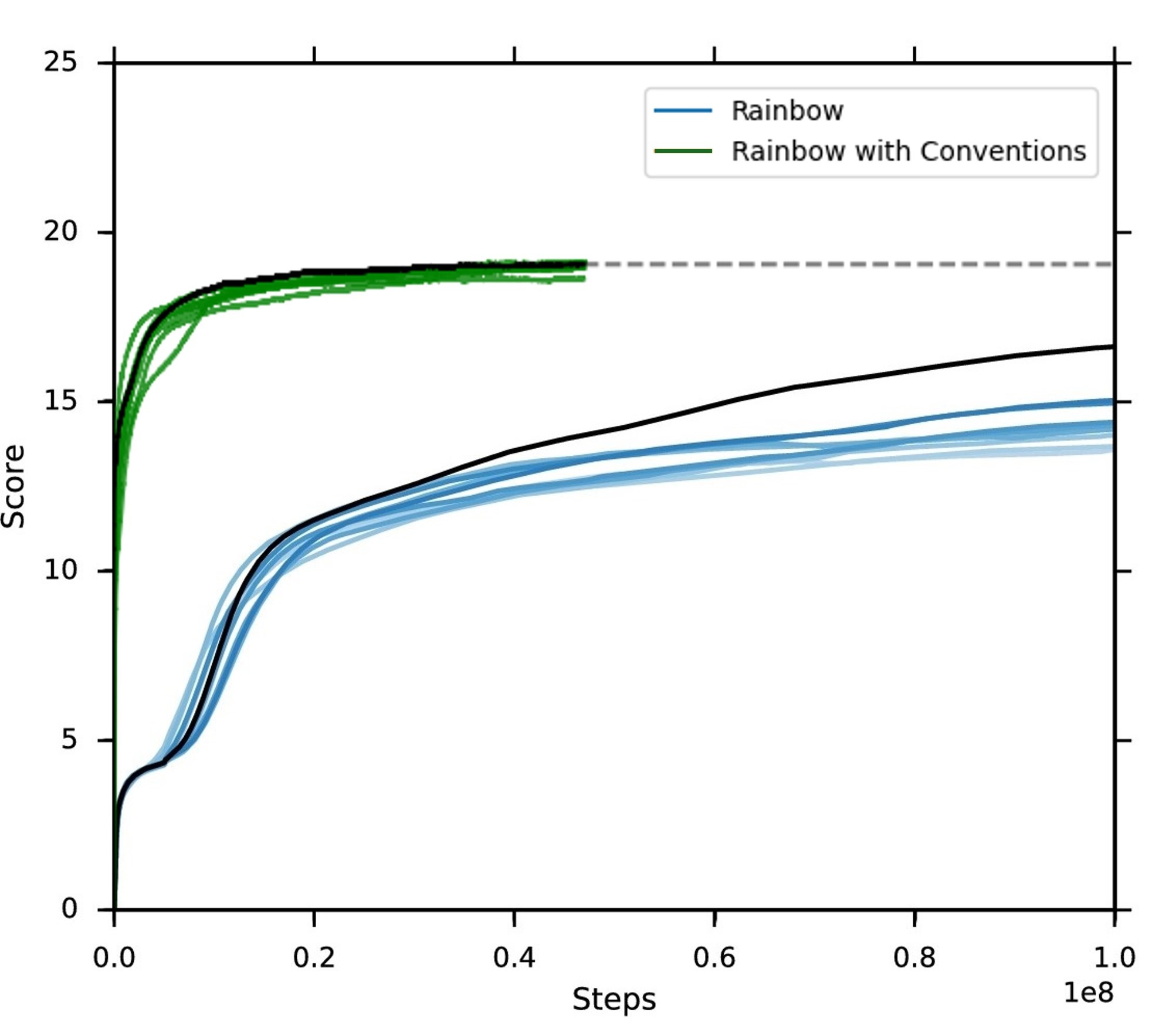}
            \caption{5-Player learning curves}\label{5p_lr}
        \end{subfigure}
        \hfil
        \caption{Learning curves with exponential moving averages (weight=0.9995) for Rainbow, obtained from Bard \textit{et al.}~\cite{hanabi_ai}, as baseline compared to Rainbow
        with an augmented action-\emph{convention} space for Hanabi two-to five-players. The best agent is highlighted in black for each agent scenario within each player
        count.}\label{lr_curves_multi}
    \end{figure}

    The learning curves for Rainbow with and without an augmented action-\emph{convention} space for 2--5 player Hanabi are shown in Fig.~\ref{lr_curves_multi}. The
    \revision{two-player} scenario seen in Fig.~\ref{2p_lr} shows the smallest improvement when applying \emph{conventions}, however the \revision{two-player} Hanabi problem can be
    considered a special case. This is due to the fact that the shared information between the two players are far more limited when compared to higher player counts. For example,
    in \revision{two-player} Hanabi, half of the hands are hidden to a single player whereas in the \revision{three-player} scenario only a third of the hands are hidden, and the
    players always have a common hand visible. This allows for more advanced reasoning over another player's actions, evident by the fact that two of the conventions, specifically
    the \emph{Finesse} and the \emph{Prompt} as discussed in Appendix~\ref{imp_convs_hanabi}, cannot be applied in \revision{two-player} Hanabi.  

\begin{figure}[!h]
        \centering
        \begin{subfigure}[b]{0.49\textwidth}
            \centering
            \includegraphics[width=\textwidth]{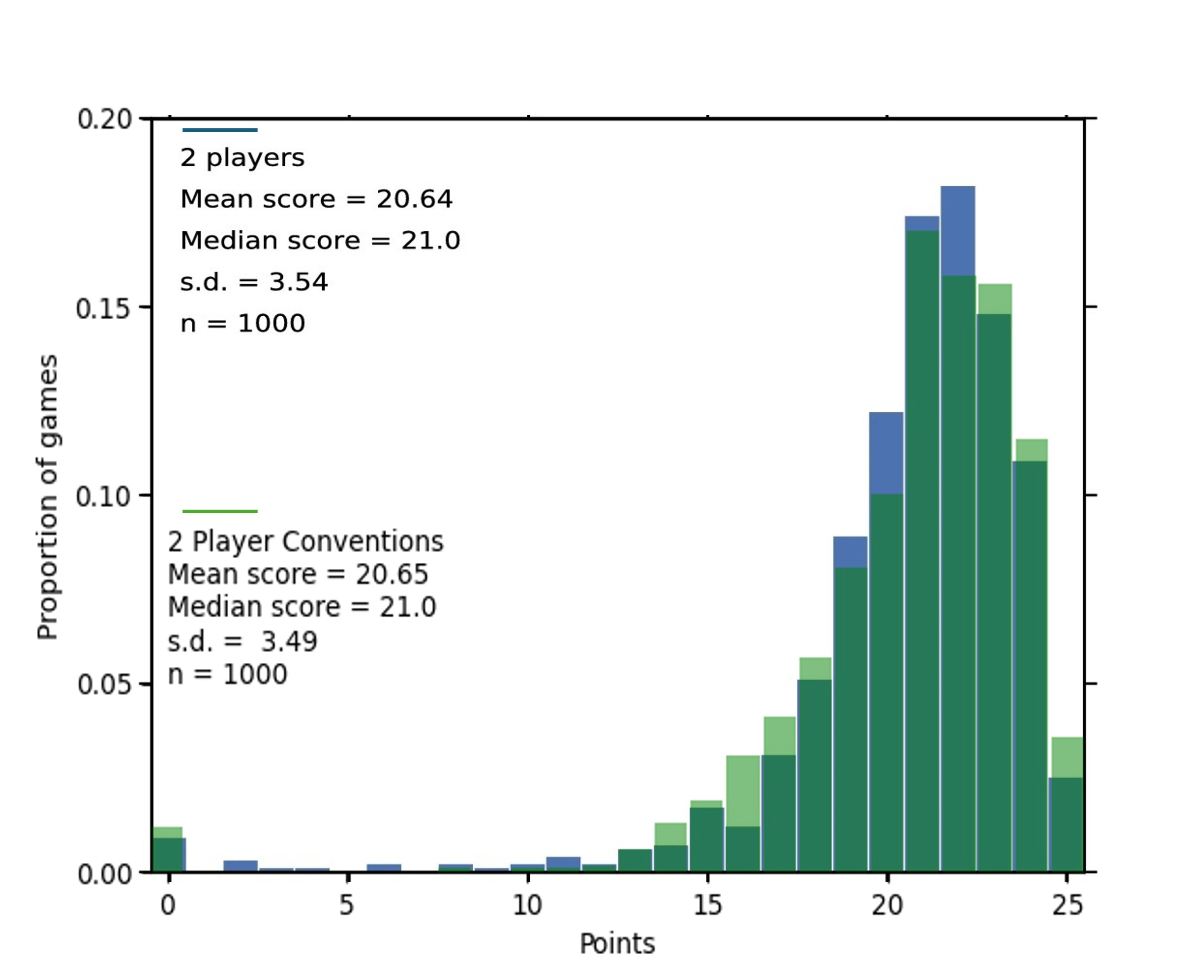}
            \caption{2-Player}\label{2p_eval_dis}
        \end{subfigure}
        \hfill
        \begin{subfigure}[b]{0.49\textwidth}
            \centering
            \includegraphics[width=\textwidth]{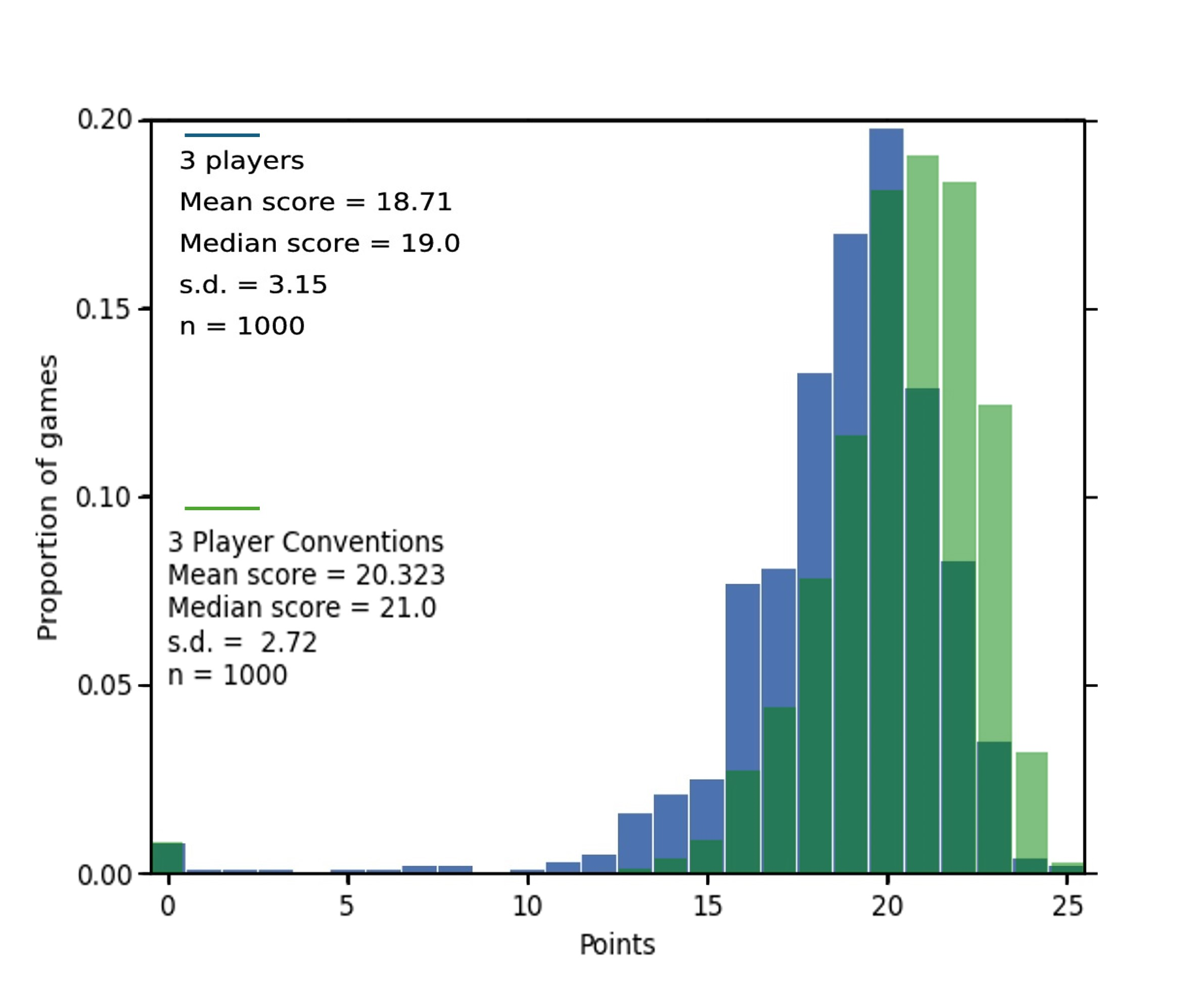}
            \caption{3-Player}\label{3p_eval_dis}
        \end{subfigure}
        \hfil
        \centering
        \begin{subfigure}[b]{0.49\textwidth}
            \centering
            \includegraphics[width=\textwidth]{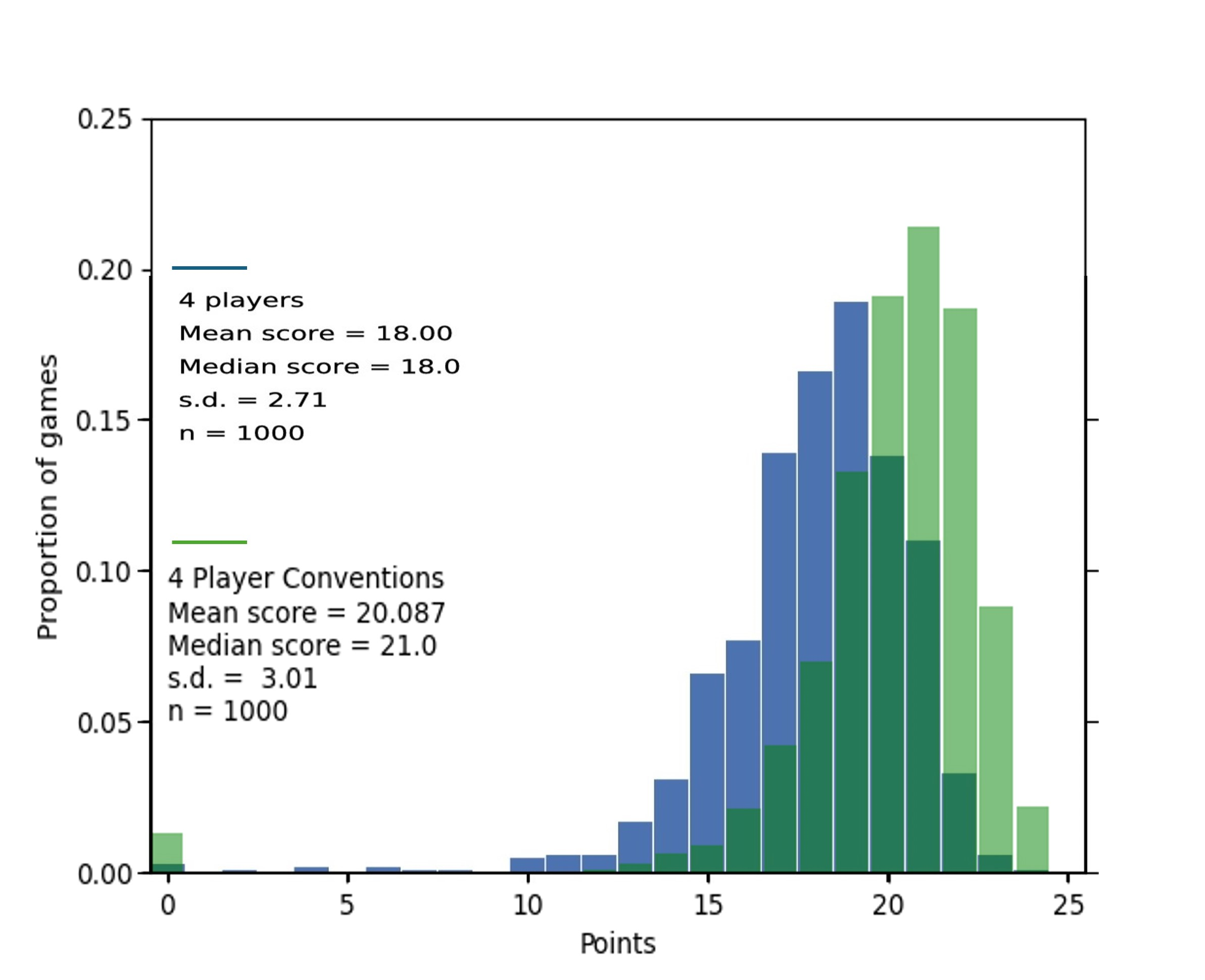}
            \caption{4-Player}\label{4p_eval_dis}
        \end{subfigure}
        \hfill
        \begin{subfigure}[b]{0.49\textwidth}
            \centering
            \includegraphics[width=\textwidth]{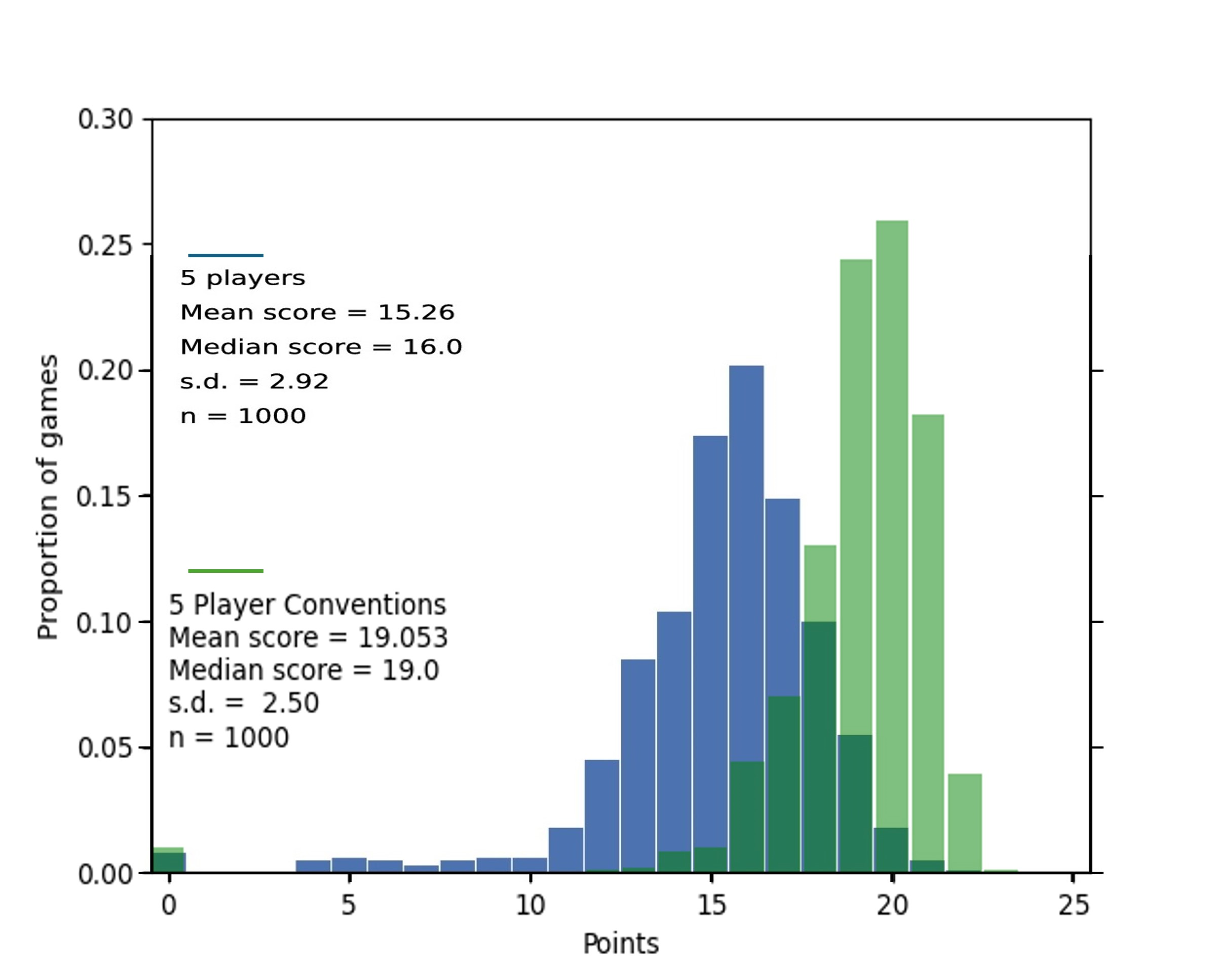}
            \caption{5-Player}\label{5p_eval_dis}
        \end{subfigure}
        \hfil
        \caption{Distribution of scores for 2--5 player Hanabi over the course of 1000 evaluation episodes comparing baseline Rainbow and Rainbow with an augmented
        action-\emph{convention} space. The baseline Rainbow results were obtained from Bard~\textit{et al.}~\cite{hanabi_ai}.}\label{eval_hists_multi}
    \end{figure}

    Despite the fact that the complexity of the problem increases exponentially as more agents are added to the environment, the performance uplift of \emph{conventions} become
    more apparent. In the \revision{three-player} setting seen in Fig.~\ref{3p_lr}, the agents are able to achieve a significantly faster training time as well as an improved
    convergent performance with a higher average score. The performance improvement becomes even more apparent when looking at the \revision{five-player} setting, seen in
    Fig.~\ref{5p_lr}, which is generally considered the most difficult problem to solve, and where the agents train roughly five times faster than the baseline Rainbow agents. In
    Fig.~\ref{eval_hists_multi} the distribution of scores during evaluation of each agent over the course of 1000 episodes is shown. In each scenario, \emph{conventions} allow for
    a significant performance improvement, achieving a consistent grouping near a score of 21/25 and less overall variance when compared to baseline Rainbow. Rainbow with an
    augmented action-\emph{convention} space also displays a robustness to increased player counts, with a remarkably consistent behaviour across all scenarios. 

    Table~\ref{eval_results} shows the average score for the best agent from each player count's various runs compared to the results from Bard \textit{et al.}~\cite{hanabi_ai}.
    Rainbow with an augmented action-\emph{convention} space demonstrates an improved score over baseline Rainbow for each scenario, with a higher average in the 3--5 player
    scenario and overall less deviation. Additionally, the \revision{three-and five-player} results are competitive compared to other algorithms, such as ACHA, while also
    requiring substantially less training steps. In the case of five-player Hanabi, Rainbow with \textit{conventions} outperform ACHA where ACHA required 20 billion training steps
    and Rainbow with \textit{conventions} only required 30 million. A statistical significance test, conducted in Appendix~\ref{stats}, shows that majority of the results are
    statistically different. However, the Rainbow with \emph{conventions} agent's scores are statistically similar to that of baseline Rainbow \revision{two-players} and ACHA
    \revision{three-players}, however \emph{conventions} allow the Rainbow algorithm to reach these scores significantly faster (5x compared to baseline Rainbow and 1000x compared
    to ACHA). Furthermore, even though these two instances are statistically similar, the agents learn completely different policies, as seen by the cross-play performance
    evaluation in the next section.

    \begin{table} [h]
        \caption{Performance evaluation over 1000 episodes for the best performing Rainbow agent with an augmented action-\emph{convention} space applied taken from a sample of 10
        different runs for each player count, compared to the results presented by Bard \textit{et al.}~\cite{hanabi_ai}. Each score is averaged out of 25 with the standard error
        of the mean shown in brackets, and the number of perfect games (25/25) shown as a percentage.}\label{eval_results}
        \begin{tabular}{@{}|p{100pt}|p{48pt}|p{48pt}|p{48pt}|p{48pt}|@{}}
            \hline
            \textbf{Method}                     &   \textbf{2P}             &   \textbf{3P}             &   \textbf{4P}             &   \textbf{5P}         \\
            \hline
            Rainbow with Conventions            &   20.65 (0.11) 3.6\%      &   20.32 (0.07) 0.2\%      &  20.09 (0.09) 0.1\%       &  19.05 (0.08) 0\%     \\
            \hline
            Rainbow~\cite{hanabi_ai}            &   20.64 (0.22) 2.5\%      &   18.71 (0.20) 0.2\%      &  18.00 (0.17) 0\%         &  15.26 (0.18) 0\%     \\
            \hline
            ACHA~\cite{hanabi_ai}               &   22.73 (0.12) 15.1\%     &   20.24 (0.15) 1.1\%      &  21.57 (0.12) 2.4\%       &  16.80 (0.13) 0\%     \\
            \hline
        \end{tabular}
    \end{table}
    
    Upon investigation, the agents tend to choose \emph{conventions} over primitive actions 70\% of the time. This is the main reason for the Rainbow agents not achieving a higher
    average score, since they struggle to learn some of the conventions available to them (specifically \emph{The Chop} as discussed in Appendix~\ref{imp_convs_hanabi}). We believe
    this to be a result of the Rainbow algorithm not having a memory of past observations, and therefore not being able to realise the importance of \emph{conventions} that have a
    high payoff over extended time frames. 
    
\subsection{Cross-play Performance}
    Although self-play has historically led to the highest state-of-the-art agent performance within Hanabi~\cite{sad_search}, cross-play is also a crucial and widely applicable
    problem to solve. The ability for agents to cooperate across different training runs or regimes, or even across architectures, can greatly benefit MARL algorithms and their
    ability to cooperate in real-world scenarios~\cite{sp_vs_cp_1,sp_vs_cp_2}. Self-play agents are notorious for not being able to cooperate with never-before-seen partners, and
    often results in very poor performance~\cite{hanabi_ai,otherplay,sp_vs_cp_1,sp_vs_cp_2}. 

    Table~\ref{eval_results_cp} shows the \revision{two-player} performance for various agents in cross-play Hanabi taken from existing literature, compared to the results of our
    cross-play Rainbow agents with an augmented action-\emph{convention} space. The results for the Rainbow with \emph{conventions} agents are statistically different from the
    other results reported in Table~\ref{eval_results_cp}, as seen by Table~\ref{stats_sig} in Appendix~\ref{stats}. \emph{Conventions} are able to significantly improve on the
    performance of cross-play agents, allowing the baseline Rainbow agents to go from abysmal performance to an acceptable score. Furthermore, it is able to outperform SAD with
    only other-play applied for the \revision{two-player} scenario, with other-play and auxiliary task applied to SAD still holding the best score. However, as mentioned
    previously, auxiliary tasks only benefit the \revision{two-player} scenario in Hanabi~\cite{sad_hanabi}. When moving from self-play to cross-play, Rainbow with an augmented
    action-\emph{convention} space only suffered a slight performance decrease and maintained remarkable consistency across various agent parings. 

    \begin{table} [h]
        \caption{Performance evaluation in \revision{two-player} Hanabi for the best performing agents in a cross-play scenario over the course of 1000 episodes. The Rainbow with
        an augmented action-\emph{convention} space agents were chosen from the top performing two agents in 10 different runs. Each score is averaged out of 25 with the standard
        error of the mean shown in brackets.}\label{eval_results_cp}
        \begin{tabular}{@{}|p{180pt}|p{60pt}|p{60pt}|@{}}
            \hline
            \multirow{2}{*}{\textbf{Method}}                            &   \multicolumn{2}{c|}{\textbf{2P}}    \\
                                                                        &   Self-play (SP)  &   Cross-play (CP) \\
            \hline 
            Rainbow with Conventions                                    &   20.65 (0.11)    &   17.02 (0.25)    \\
            \hline
            Rainbow~\cite{hanabi_ai}                                    &   20.64 (0.22)    &   2.91 (1.67)     \\
            \hline
            ACHA~\cite{hanabi_ai}                                       &   22.73 (0.12)    &   3.31 (1.78)     \\
            \hline
            SAD~\cite{otherplay}                                        &   23.94 (0.04)    &   2.52 (0.34)     \\
            \hline
            SAD with other-play~\cite{otherplay}                        &   23.93 (0.02)    &   15.32 (0.65)    \\
            \hline
            SAD with auxiliary tasks and other-play~\cite{otherplay}    &   24.06 (0.02)    &   22.07 (0.11)    \\
            \hline
        \end{tabular}
    \end{table}

    Table~\ref{eval_results_cp_all} shows the results for cross-play Rainbow with an augmented action-\emph{convention} space for all player counts of Hanabi. As seen in the
    self-play results for Rainbow with \emph{conventions} and confirmed in Table~\ref{eval_results_cp_all}, the \revision{two-player} scenario is a special case, with the agents
    showing the biggest performance degradation when moving from self-play to cross-play, and overall worst score compared to the other cross-play results. While most research
    efforts on cross-play Hanabi focus on the \revision{two-player} scenario, our results demonstrate that more powerful \emph{conventions} are applicable in higher player counts,
    thus, it is important to \revision{consider} all the player scenarios when evaluating cross-play performance. 

    \begin{table} [h]
        \caption{Performance evaluation in 2--5 players Hanabi for the best performing Rainbow with an augmented action-\emph{convention} space agents in a cross-play scenario over
        the course of 1000 episodes. Each combination of agents in each player count were chosen from the top performing agents in a sample of 10 different runs. Each score is
        averaged out of 25 with the standard error of the mean shown in brackets.}\label{eval_results_cp_all}
        \begin{tabular}{@{}|p{80pt}|p{20pt}|p{20pt}|p{20pt}|p{20pt}|p{20pt}|p{20pt}|p{20pt}|p{20pt}|@{}}
            \hline
            \multirow{2}{*}{\textbf{Method}}        &   \multicolumn{2}{c}{\textbf{2P}}     &   \multicolumn{2}{c}{\textbf{3P}}     &   \multicolumn{2}{c}{\textbf{4P}}     &  \multicolumn{2}{c|}{\textbf{5P}}     \\
                                                    &   SP              &     CP            &   SP              &   CP              &   SP              &       CP          &   SP              &   CP                       \\
            \hline 
            Rainbow with Conventions                &   20.65 (0.11)    &   17.02 (0.25)    &   20.32 (0.07)    &   18.60 (0.15)    &  20.09 (0.09)     &   18.56 (0.11)    &   19.05 (0.08)    &   17.69 (0.09)    \\
            \hline
        \end{tabular}
    \end{table}

    Most remarkable is the \revision{five-player} results, where the cross-play Rainbow agents with an augmented action-\emph{convention} space are able to achieve a relatively
    high score and still cooperate effectively. The reason \emph{conventions} are able to offer such a large performance gain for cross-play is due to the fact that the agents
    learn from a communal list of conventions that are taken from existing domain knowledge. Thus, if an agent is paired with a never-before-seen cooperator that has learned to
    play with the same list of conventions, they will still be able to cooperate effectively, since there is no uncertainty over another agent's intent and reasoning. This is not
    unexpected, since humans who learn from the same ``list'' of conventions (whether it is physical or passed down), and that have never before interacted, can cooperate and
    communicate effectively with relatively low uncertainty. 

\section{Conclusion}\label{concl} 
    Hanabi offers a complex and intricate problem for multi-agent reinforcement learning agents, since it incorporates various features of real-world problems, such as those found
    in the autonomous agents controlling vehicles problem. It tests an agent's ability to reason over another player's actions and intensions, while maintaining a limited
    communication channel and partial observability. Existing MARL algorithms focus on complex architecture design and implement advanced algorithms capable of achieving remarkable
    performance. However, these algorithms are often computationally expensive, and require immense amounts of training data to learn convergent policies. In this paper, we focused
    on a different approach, exploring another core aspect of the Hanabi problem, namely conventions, and how to incorporate them into MARL. We showed how existing human
    conventions can be implemented in a MARL scenario using a special form of cooperative actions and action space augmentations. 
    
    Our main results show that \emph{conventions} are able to significantly improve on the training time for Rainbow agents in Hanabi, and in the case of three-to five-players,
    also improve on the converged policy. Other Hanabi algorithms, such as SAD or MAPPO, require billions of training steps to reach convergent policies, whereas \emph{conventions}
    reduced the number of training steps for Rainbow to below 30 million consistently. Additionally, \emph{conventions} showed significant performance increase for cross-play
    agents, and allowed the Rainbow agents to go from not cooperating at all to being able to achieve decent performance across all player scenarios. For the most difficult
    scenario of five players, the cross-play Rainbow \emph{conventions} agents trained on 50 million steps are able to outperform the ACHA self-play agents trained on 20 billion
    steps. Our work, as presented in this paper, mainly focused on single-and two-step \emph{conventions} in the Hanabi problem, with the goal of extending these concepts to other
    multi-agent scenarios (including simultaneous action spaces) containing multistep \emph{conventions}, as well as the exploration of more advance conventions (and the
    application of those conventions) on more advanced algorithms within the Hanabi setting.

    An additional potential benefit of \textit{conventions} is that of encrypted communication, where a malicious user tries to intercept the ad-hoc agent's communications, for
    example military communication or team communication in sports. \textit{Conventions} can allow for an additional means of communicating encrypted sensitive data, even if the
    main communication channel is compromised, since implicit communication takes place through the observed actions, and can be explored in future research. However, we believe
    the biggest avenue for future research is exploring and developing \emph{convention} discovery, similar to that of option discovery, which will allow agents to produce and
    define their own \emph{conventions} as they train and learn. We believe that our research will serve as the foundation for this future endeavour, proving the importance of
    conventions within MARL problems containing partial observability and limited communication. 

\backmatter

\bmhead{Acknowledgements}

This research was funded in part by the DW Ackerman Bursary as distributed by Stellenbosch University. 

\bmhead{Author contributions}

All authors contributed to the study conception and design. Material preparation and data collection were performed by F. Bredell, and all authors contributed equally to analysing
the data. The first draft of the manuscript was written by F. Bredell and all authors commented on previous versions of the manuscript. All authors read and approved the final
manuscript.

\section*{Declarations}
\begin{itemize}
\item \textbf{Conflict of interest}                         The authors declare no competing interests.
\item \textbf{Ethics approval and consent to participate}   Not applicable.
% \item Consent for publication
% \item Materials availability
\item \textbf{Code availability}                            Code for this research is publicly available at \url{https://github.com/FBredell/MARL_artificial_conventions_Hanabi}.
\end{itemize}

% \noindent
% If any of the sections are not relevant to your manuscript, please include the heading and write `Not applicable' for that section. 

% %%===================================================%%
% %% For presentation purpose, we have included        %%
% %% \bigskip command. Please ignore this.             %%
% %%===================================================%%
% \bigskip

% Springer journals and proceedings: \url{https://www.springer.com/gp/editorial-policies}

% Nature Portfolio journals: \url{https://www.nature.com/nature-research/editorial-policies}

% \textit{Scientific Reports}: \url{https://www.nature.com/srep/journal-policies/editorial-policies}

% BMC journals: \url{https://www.biomedcentral.com/getpublished/editorial-policies}
% \end{flushleft}

\begin{appendices}
% \newpage
\section{Implemented Conventions and Principles}\label{imp_convs}
    Herein follows a list of the implemented \textit{conventions} and the principles upon which they are built for Small Hanabi and Full Hanabi. 

\subsection{Small Hanabi}\label{imp_convs_small_hanabi}
    The principles and conventions for Small Hanabi were developed in-house through multiple human plays and, therefore, we do not claim for them to be the most optimal. However,
    it served as a good starting point for investigating the capabilities of \textit{conventions}. Furthermore, these conventions allow for a pure \textit{convention} space, i.e.
    the \textit{conventions} account for all possible states and an agent will always have a \text{convention} available to take. Note that any hint convention is only available to
    take if there is available hint tokens to spend, as per the rules of the game.
    \hfil \\
    \noindent \textbf{Principles:}
    \begin{enumerate}
        \item Colour saves -- Players use colour hints to indicate to the other player that the card must not be discarded.
        \item Value plays -- Players use value hints to indicate to the other player that a card is playable, unless the value hinted card is clearly not playable (i.e. both stacks are
        currently above the hinted value), then the card is discardable.
        \item Pseudo-chop -- Players always discard oldest non-colour hinted card, unless both cards are colour hinted and a player must discard, then that player discards their oldest card. 
        \item Better to let go -- If the current score is less than 7, a player can reveal a five in another player's hand to indicate to that player that they must discard that five.
        \item Double information -- If both cards share information (i.e. colour or rank), a player can hint to another player the shared information to indicate that the newest
        card is playable. 
        \item Implied moves -- If the negative information produced by other hints indicates a card as playable or discardable, a player can play or discard that card. 
        \item We need hints -- Players should prioritise discarding for hint tokens rather than playing, especially if there is only one hint token remaining. 
    \end{enumerate}
    \hfil \\
    \indent These principles lead to a set of \textit{conventions}, shown in Table~\ref{small_hanabi_conv}, and result in policies allowing an agent to follow their "rules". In
    practice each convention with its set of conditions $\lambda_k$ will form part of an agent's actions in their action space. Therefore, if the agent has a pure
    \textit{convention} space with the \textit{conventions} in Table~\ref{small_hanabi_conv}, the size of the \emph{convention}-step space will be $|C|=12$. Each
    \textit{convention}, and its subsequent conditions, will become available to an agent through the use of action masking based on the current observation, as described in
    Section~\ref{aug}.

    \begin{table} [!h]
        \caption{Table detailing all the implemented \textit{conventions} in Small Hanabi based on specific principles. Each \emph{convention} $c_k$ has a step-size of $m_k$, and
        subsequently contains a set of conditions $\lambda_k^{0:m_k}$ corresponding to the set of policies $\pi_k^{0:m_k}$ that determine the appropriate environment action $A_t$.
        However, single-step \textit{conventions} only contain one condition and one policy, i.e. $m_k=1$, which is simultaneously the initial and final condition and policy of
        that \textit{convention}. }\label{small_hanabi_conv}
        \centering
        \begin{tabular}{@{}|p{6pt}|p{8pt}|p{10pt}|p{10pt}|p{250pt}|p{40pt}|@{}}
            \hline
            \mystrut \textbf{$k$}  & \textbf{$m_k$} & \textbf{$\lambda_k^m$} & \textbf{$\pi_k^m$} & \textbf{Policy Description} & \textbf{Principles} \\
            \hline
            \mystrut $0$        & $1$ & $\lambda_0^1$ & $\pi_0^1$ & Discard oldest non-colour hinted card. If both are colour hinted, then discard the oldest card   & 3, 7   \\
            \hline 
            \mystrut $1$        & $2$ & $\lambda_1^1$ & $\pi_1^1$ & Give a value hint for a playable card in the other player's hand  & 2   \\
            \hline
            \mystrut $1$        & $2$ & $\lambda_1^2$ & $\pi_1^2$ & Play a value hinted card received from the other player  & 2  \\
            \hline
            \mystrut $2$        & $2$ & $\lambda_2^1$ & $\pi_2^1$ & If the score is less than 7, and the other player has a five of any colour, value hint that five to indicate to that player that they should discard the five & 4   \\
            \hline
            \mystrut $2$        & $2$ & $\lambda_2^2$ & $\pi_2^2$ & If the score is less than 7, and a value hinted five (which is clearly not playable) was received from the other player, discard the value hinted five & 4   \\
            \hline
            \mystrut $3$        & $1$ & $\lambda_3^1$ & $\pi_3^1$ & If a card is critical, i.e. it must not be discarded, colour hint to the other player so save the card from being discarded  & 1, 3   \\
            \hline
            \mystrut $4$        & $2$ & $\lambda_4^1$ & $\pi_4^1$ & If a card is discardable, value hint that card to the other player to indicate that they can discard it  & 2, 7   \\
            \hline
            \mystrut $4$        & $2$ & $\lambda_4^2$ & $\pi_4^2$ & If an individual card was value hinted by the other player, and is clearly not playable, discard that card & 2, 7   \\
            \hline
            \mystrut $5$        & $2$ & $\lambda_5^1$ & $\pi_5^1$ & If the newest card in the other player's hand is playable, and their two card share information (colour or rank), hint the shared information  & 5   \\
            \hline
            \mystrut $5$        & $2$ & $\lambda_5^2$ & $\pi_5^2$ & If the other player has hinted information about both cards in hand, the newest one is playable and can, therefore, be played  & 5   \\
            \hline
            \mystrut $6$        & $1$ & $\lambda_6^1$ & $\pi_6^1$ & If the implied knowledge reveals a card as playable, then a player can play that card  & 6   \\
            \hline
            \mystrut $7$        & $1$ & $\lambda_7^1$ & $\pi_7^1$ & If the oldest card in hand is value hinted and can be played in the future (achieved through convention 5) and hint tokens are $<=$ 1, then a player
            should discard their newest card (override pseudo-chop)  & 3, 5, 7   \\
            \hline
        \end{tabular}
    \end{table}
% \newpage
\subsection{Hanabi}\label{imp_convs_hanabi} 
    The principles for Hanabi are based on those defined by the \href{https://hanabi.github.io/}{H-Group}. We strongly encourage the reader to read their documentation,
    specifically chapters \RomanNumeralCaps{1} -- \RomanNumeralCaps{8}, on Hanabi, conventions, and convention principles for an in-depth discussion (with examples) for each
    included principle and their corresponding conventions discussed below. As discussed in Section~\ref{small_hanabi_results}, the implemented \textit{conventions} only consist of
    the most fundamental principles, since these are generally considered standardised within the Hanabi community. Majority of the listed \textit{conventions} are single-or
    two-step \textit{conventions}, however the \textit{Prompt} and the \textit{Finesse} are multistep. Additionally, the \href{https://hanabi.github.io/}{H-Group} discuss more
    advance multistep conventions useful for future considerations. Note that the prompt and finesse is only applicable in three-to five-player Hanabi, however there does exist a
    self-prompt for two-player Hanabi which reduces the \emph{convention} to a two-step \emph{convention}, and is included in our implementation. \hfil \\
    \noindent \textbf{Principles:}
    \begin{enumerate}[nolistsep]
        \item The Chop -- A player should discard their oldest non-hinted card, called ``the chop''. If all cards are hinted and not obviously discardable, then no cards should be
        discarded.
        \item Save Hints -- A player must try to prevent another player from discarding an important card when it is on the chop by hinting the card's colour or value.
        \item Play Hints -- A player must try to promise another player that a card is playable. If a hint is obviously playable (i.e. not on the chop or a value hint on the chop
        that follows a current stack), then it is a play hint. However, if a hint touches the chop, and it is unclear if it was a play or save hint, then assume it was a save hint. 
        \item Single card focus -- Any hint given or received should always have one card as the focus. This principle is divided into three scenarios:
        \begin{enumerate}
            \item If only one card is hinted, then that card is the focus. 
            \item If two or more cards are hinted and one is on the chop, then the focus was on the chop (either play or save).
            \item If two or more cards are hinted and none are on the chop, then the focus was on the newest card (play).
        \end{enumerate}
        \item Good touch principle -- A player should only give information on cards that will be playable. A player should not give hints for cards that will never be played,
        ensuring that they will become ``old'' and become the chop (to get discarded).
        \item Save principle -- A player must try to save all critical cards if these cards are on the chop position. Critical cards include:
        \begin{enumerate}
            \item Any five.
            \item A card is the last of its kind, i.e. there is no more of that card in the deck based on the discard pile.
            \item Any unique two, based on the current player's perspective, i.e. none of the other players have the same coloured two in hand. 
        \end{enumerate}
        \item Minimum clue value principle -- Every hint should include at least one non-hinted card.
        \item Prompts -- If the next two players have cards following on each other in order of play, i.e. the first player has the first card and the second player has the second
        card following that card, with the first player having partial knowledge of their card (usually colour), then the current player can give a play clue to the second player.
        The first player will see this clue, realise it is a play clue on a card that is not yet playable, and deduce that their partially revealed card must be the card required
        to make the play clue valid, otherwise the second player will make a mistake. 
        \item Finesse -- Similar to prompt, however the first player's card is in the newest position \textit{and} they have no knowledge of that card. This forces a blind play,
        since the first player will see a false play clue given to the second player, thereby realising they must have the required card to complete the play, and since none of
        their cards have partial information regarding the sequence (i.e. it is not a prompt), their newest card must be the required card.
    \end{enumerate}
    \hfil \\
    There also exists implied principles, which are not directly translated into \textit{conventions}, however they do govern some of the internal reasoning of certain
    \textit{conventions}. These implied principles include:
    \hfil \\
    \begin{enumerate}[resume]
        \item Perfect teammates -- A player should assume that their teammates will never make a mistake and any clue given has meaning.
        \item Colour over value hints -- In general, colour hints give more information than value hints and should be the preferred hint. However, there does exist situations
        where value hints are more beneficial and should, therefore, be used. 
        \item Early game -- At the start of the game, if there are other \textit{conventions} available to a player, they should not discard their chop. As soon as the first card
        is discarded, i.e. a player had no other option, then the chop discard becomes available to all players.
    \end{enumerate}
    \hfil \\
    \indent These principles and implied principles lead to the necessary set of \textit{conventions}, shown in Table~\ref{hanabi_conv}, and result in policies allowing an agent to
    follow their "rules". Similar to the Small Hanabi \textit{conventions}, each \textit{convention} with its set of conditions will act as an action in the agent's action space.
    If the agent has a pure \textit{convention} space with the \textit{conventions} in Table~\ref{hanabi_conv}, the \emph{convention}-step space will have a size of $|C|=21$.
    However, unlike Small Hanabi, these \textit{conventions} do not account for each possible game state, therefore, an augmented \textit{convention}-action space is required, as
    discussed in Section~\ref{aug}. 

    \begin{table}[!h]
        \caption{Table detailing all the implemented \textit{conventions} in Hanabi based on discussed principles. Each \emph{convention} $c_k$ has a step-size of $m_k$, and
        subsequently contains a set of conditions $\lambda_k^{0:m_k}$ corresponding to a set of policies $\pi_k^{0:m_k}$ that determine the appropriate environment action $A_t$.
        For simplicity, this table shows the \textit{conventions} applicable to player 1 in a three-player scenario, and as the player count increases or decreases, certain
        \textit{conventions} also increase or decrease to account for different player counts. }\label{hanabi_conv}
        \centering
        \begin{tabular}{@{}|p{6pt}|p{8pt}|p{10pt}|p{10pt}|p{250pt}|p{40pt}|@{}}
            \hline
            \mystrut \textbf{$k$}  & \textbf{$m_k$} & \textbf{$\lambda_k^m$} & \textbf{$\pi_k^m$} & \textbf{Policy Description} & \textbf{Principles} \\
            \hline
            \mystrut $0$           & $2$ & $\lambda_0^1$ & $\pi_0^1$ & Give a value hint for a playable card in player 2's hand (next player)  & 3, 4, 5, 10, 11   \\
            \hline
            \mystrut $0$           & $2$ & $\lambda_0^2$ & $\pi_0^2$ & Play value hinted card in hand received from player 3 (previous player)  & 3, 4, 5, 10, 11   \\
            \hline
            \mystrut $1$           & $2$ & $\lambda_1^1$ & $\pi_1^1$ & Give a value hint for a playable card in player 3's hand   & 3, 4, 5, 10, 11   \\
            \hline
            \mystrut $1$           & $2$ & $\lambda_1^2$ & $\pi_1^2$ & Play value hinted card in hand received from player 2  & 3, 4, 5, 10, 11   \\
            \hline
            \mystrut $2$           & $2$ & $\lambda_2^1$ & $\pi_2^1$ & Give a colour hint for a playable card in player 2's hand (next player)   & 3, 4, 5, 10, 11   \\
            \hline
            \mystrut $2$           & $2$ & $\lambda_2^2$ & $\pi_2^2$ & Play colour hinted card in hand received from player 3 (previous player)   & 3, 4, 5, 10, 11   \\
            \hline
            \mystrut $3$           & $2$ & $\lambda_3^1$ & $\pi_3^1$ & Give a colour hint for a playable card in player 3's hand   & 3, 4, 5, 10, 11   \\
            \hline
            \mystrut $3$           & $2$ & $\lambda_3^2$ & $\pi_3^2$ & Play colour hinted card in hand received from player 2   & 3, 4, 5, 8, 9, 10, 11   \\
            \hline
            \mystrut $4$           & $1$ & $\lambda_4^1$ & $\pi_4^1$ & If player 2's chop is a 5, then value hint to save that card from being discarded   & 1, 2, 6(a), 7   \\
            \hline
            \mystrut $5$           & $1$ & $\lambda_5^1$ & $\pi_5^1$ & If player 3's chop is a 5, then value hint to save that card from being discarded   & 1, 2, 6(a), 7   \\
            \hline
            \mystrut $6$           & $1$ & $\lambda_6^1$ & $\pi_6^1$ & If player 2's chop is a unique 2, then value hint to save that card from being discarded & 1, 2, 5, 6(b), 7   \\
            \hline
            \mystrut $7$           & $1$ & $\lambda_7^1$ & $\pi_7^1$ & If player 3's chop is a unique 2, then value hint to save that card from being discarded   & 1, 2, 5, 6(b), 7   \\
            \hline
            \mystrut $8$           & $1$ & $\lambda_8^1$ & $\pi_8^1$ & If player 2's chop is a critical card, i.e. if that card is discarded there will not be another one for the rest of the game, colour or value hint
            to save that card from being discarded, based on the other cards in that player's hand and the other principles   & 1, 2, 5, 6(c), 7, 11   \\
            \hline
            \mystrut $9$           & $1$ & $\lambda_9^1$ & $\pi_9^1$ & If player 3's chop is a critical card, colour or value hint to save that card from being discarded, based on the other cards in that player's hand
            and the other principles & 1, 2, 5, 6(c), 7, 11   \\
            \hline
            \mystrut $10$          & $3$ & $\lambda_{10}^1$ & $\pi_{10}^1$ & Prompt player 2 to play a partially revealed card in their hand, if player 3 has the card following that card. This is achieved by a clear play-hint to
            player 3, which player 2 will perceive and recognise as an ``incorrect'' hint, unless their partially revealed card is the one required to make it correct  & 3, 4, 5,
            7, 8, 10, 11    \\
            \hline
            \mystrut $10$          & $3$ & $\lambda_{10}^2$ & $\pi_{10}^2$ & React to a prompt started by player 3, who gave a clear play-hint to player 2, however that card is not playable. Therefore, the partially
            revealed card in hand must be playable  & 3, 4, 5, 7, 8, 10, 11    \\
            \hline
            \mystrut $10$          & $3$ & $\lambda_{10}^3$ & $\pi_{10}^3$ & Finish prompt started by the player before the previous player, with a similar structure as \emph{convention} $k=3$ & 3, 4, 5, 8, 9, 10, 11 \\
            \hline
        \end{tabular}
    \end{table}

    \begin{table} [!h]
        \caption*{\textbf{Table~\ref{hanabi_conv} Continued}}\label{hanabi_conv_cont}
        \centering
        \begin{tabular}{@{}|p{6pt}|p{8pt}|p{10pt}|p{10pt}|p{250pt}|p{40pt}|@{}}
            \hline
            \mystrut \textbf{$k$}  & \textbf{$m_k$} & \textbf{$\lambda_k^m$} & \textbf{$\pi_k^m$} & \textbf{Policy Description} & \textbf{Principles} \\
            \hline
            \mystrut $11$          & $3$ & $\lambda_{11}^1$ & $\pi_{11}^1$ & Finesse player 2 to play their newest card in hand, if player 3 has the card following that card. This is achieved by a clear play-hint to player 3, which
            player 2 will perceive and recognise as an ``incorrect'' hint, and they have no partially revealed cards which would indicate a prompt, therefore their newest card must
            be the required card  & 3, 4, 5, 7, 9, 10, 11    \\
            \hline
            \mystrut $11$          & $3$ & $\lambda_{11}^2$ & $\pi_{11}^2$ & React to a finesse started by player 3, who gave a clear play-hint to player 2, however that card is not playable and there is no partially
            revealed card in hand that could indicate a prompt. Therefore, the newest card in hand must be playable  & 3, 4, 5, 7, 9, 10, 11    \\
            \hline
            \mystrut $11$          & $3$ & $\lambda_{11}^3$ & $\pi_{11}^3$ & Finish finesse started by the player before the previous player, with a similar structure as \emph{convention} $k=3$ & 3, 4, 5, 8, 9, 10, 11 \\
            \hline
            \mystrut $12$          & $1$ & $\lambda_{12}^1$ & $\pi_{12}^1$ & If there is a card in the chop position, then discard that card    & 1, 12 \\
            \hline
        \end{tabular}
    \end{table}

    As seen by the \emph{conventions} in Table~\ref{hanabi_conv}, the \textit{Prompt} and the \textit{Finesse} are the only implemented multistep \emph{conventions} and, therefore,
    the only \emph{conventions} to have a step-size $m_k > 2$. Notably, their finalising conditions $\lambda_{10}^3$ and $\lambda_{11}^3$ are identical to that of \emph{convention}
    $3$, i.e. $\lambda_{3}^2$. The reason for this is that the environment action resulting from $\pi_{10}^1$ and/or $\pi_{11}^1$ is always a colour hint to the player after the
    next player (player 3 in the case of Table~\ref{hanabi_conv}). Therefore, to complete \emph{convention} $10$ or $11$, the finalising condition and policy is identical to those
    of \emph{convention} 3, since from player's 3's perspective, they just received a clear play colour hint from the player before the previous player (for player 1 this would be
    player 2 as depicted in Table~\ref{hanabi_conv}). Thus, to simplify our implementation, in practice we combined $\lambda_{10}^3$, $\lambda_{11}^3$, and $\lambda_{3}^2$, as well
    as $\pi_{10}^3$, $\pi_{11}^3$, and $\pi_{3}^2$.
    
\newpage
\section{Hyperparameters}\label{hyperparams} 
    Herein follows a list of the hyperparameters used for independent deep Q-learning, Rainbow, and their \emph{conventions} variants shown in Table~\ref{table of hypers 2}. Each
    method's hyperparameters were optimized using parameter sweeps, i.e., training each method with various combinations of hyperparameters over the course of multiple runs and
    selecting the hyperparameters with the best results.

    \begin{table} [!h]
        \caption{Table of hyperparameter values for each method used in our evaluation of Small Hanabi and Full Hanabi 2--5 Players. The hyperparameters used for Rainbow were
        sourced from Bard \textit{et al.}~\cite{hanabi_ai}, and after testing found to be the same after implementing \emph{conventions}.}\label{table of hypers 2}
        \centering
        \begin{tabular}{@{}|p{100pt}|p{30pt}|p{250pt}|@{}}
            \hline
            \multicolumn{3}{|c|}{\textbf{Deep Q-learning (With and without \emph{conventions})}} \\
            \hline
            Hyperparameter                      &   Value   &   Description \\
            \hline
            Learning rate $\alpha$              &   0.00005 &   Learning rate used by Adam optimizer \\
            Discount factor $\gamma$            &   0.5     &   Discount factor used in Q-learning update step \\
            Training exploration $\epsilon_t$   &   0.001   &   Value of $\epsilon$ in $\epsilon$-greedy strategy during training \\
            Evaluation exploration $\epsilon_e$ &   0.00    &   Value of $\epsilon$ in $\epsilon$-greedy strategy during evaluation\\
            Training exploration decay          &   1000    &   Decay rate of $\epsilon$ starting at 1 and ending in $\epsilon_t$ \\
            Replay memory size                  &   35000   &   Where experiences are stored to be used in the update step \\
            Sampling batch size                 &   64      &   Number of experiences randomly sampled from the experience replay memory used in the update step \\
            Target network update frequency     &   200     &   Number of steps before the target network is updated with the policy network \\
            \hline
            \multicolumn{3}{|c|}{\textbf{Rainbow (With and without \emph{conventions})}} \\
            \hline
            Hyperparameter                      &   Value   &   Description \\
            \hline
            Learning rate $\alpha$              &   0.000025&   Learning rate used by Adam optimizer \\
            Discount factor $\gamma$            &   0.99    &   Discount factor used in Q-learning update step \\
            Training exploration $\epsilon_t$   &   0.00    &   Value of $\epsilon$ in $\epsilon$-greedy strategy during training \\
            Evaluation exploration $\epsilon_e$ &   0.00    &   Value of $\epsilon$ in $\epsilon$-greedy strategy during evaluation\\
            Training exploration decay          &   1000    &   Decay rate of $\epsilon$ starting at 1 and ending in $\epsilon_t$ \\
            Replay memory size                  &   50000   &   Where experiences are stored to be used in the update step \\
            Sampling batch size                 &   32      &   Number of experiences randomly sampled from the experience replay memory used in the update step \\
            Target network update frequency     &   500     &   Number of steps before the target network is updated with the policy network \\
            Distribution atoms                  &   51      &   Number of distributional atoms over which the value distributions are approximated as a discrete distributions \\
            Step count $n$                      &   1       &   Number of steps over which the return is constructed in n-step bootstrapping \\
            \hline
        \end{tabular}\label{tab6}
    \end{table} 
\newpage

\section{Statistical Significance Test}\label{stats}
    Herein follows the results for a statistical significant test comparing our results for the Rainbow with \textit{conventions} agents in self-play and cross-play against those
    of existing literature. For these test we use the unpaired Welch's t-test~\cite{welch_t_test}, and assume a result to be statistically different from another if the calculated
    \textit{p}-value is less than 0.05. We state our null hypothesis as: ``The results obtained for the performance of the Rainbow with \textit{conventions} algorithm in self-play
    and cross-play Hanabi is not statistically different to that of existing research on the Hanabi problem.''

    \begin{table}[!h]
        \caption{Table showing the results of a statistical significance test comparing the scores of Rainbow with \textit{conventions} agents in self-play and cross-play Hanabi
        against those of existing research. If a result is statistically different, i.e. a \textit{p}-value of less than 0.05 is obtained, the cell is coloured in green, and if a
        result is statistically similar it is coloured in red. }
        \centering
        \begin{tabular}{@{}|p{55pt}|p{22pt}|p{35pt}|p{22pt}|p{35pt}|p{22pt}|p{35pt}|p{22pt}|p{35pt}|@{}}
            \hline
            \textbf{Method}                     &   \textbf{2P}         & \textbf{\textit{p}-value} &   \textbf{3P}         &  \textbf{\textit{p}-value}    &   \textbf{4P}         &  \textbf{\textit{p}-value}    &   \textbf{5P}         &  \textbf{\textit{p}-value}       \\
            \hline
            \multicolumn{9}{|c|}{Self-play} \\
            \hline
            Rainbow with Conventions            &   20.65 (0.11) 3.6\%  & --                        &   20.32 (0.07) 0.2\%  & --                            &  20.09 (0.09) 0.1\%   & --                            &  19.05 (0.08) 0\%     & --   \\
            \hline
            Rainbow~\cite{hanabi_ai}            &   20.64 (0.22) 2.5\%  & \cellcolor{red!50}0.9676  &   18.71 (0.20) 0.2\%  & \cellcolor{green!50}5.906e-14 &  18.00 (0.17) 0\%     & \cellcolor{green!50}$\approx$ 0 &  15.26 (0.18) 0\%   & \cellcolor{green!50}$\approx$ 0   \\
            \hline
            ACHA~\cite{hanabi_ai}               &   22.73 (0.12) 15.1\% & \cellcolor{green!50}5.524e-36 &   20.24 (0.15) 1.1\% & \cellcolor{red!50}0.629    &  21.57 (0.12) 2.4\%   & \cellcolor{green!50}2.047e-22 &  16.80 (0.13) 0\%     & \cellcolor{green!50}$\approx$ 0   \\
            \hline
            \multicolumn{9}{|c|}{Cross-play} \\
            \hline
            Rainbow with Conventions            &   17.02 (0.25)        & --                        &   18.60 (0.15)        & --                            &   18.56 (0.11)        & --                            &   17.69 (0.09)        & --    \\
            \hline
            Rainbow~\cite{hanabi_ai}            &   2.91 (1.67)         & \cellcolor{green!50}2.22e-16 & --                 & --                            & --                    & --                            & --                    & --   \\
            \hline
            ACHA~\cite{hanabi_ai}               &   3.31 (1.78)         & \cellcolor{green!50}5.396e-14 & --                 & --                            & --                    & --                            & --                    & --   \\
            \hline
            SAD~\cite{otherplay}                &   2.52 (0.34)         & \cellcolor{green!50}$\approx$ 0 & --                 & --                            & --                    & --                            & --                    & --   \\
            \hline
            SAD with other-play~\cite{otherplay}&   15.32 (0.65)        & \cellcolor{green!50}0.01478   & --                 & --                            & --                    & --                            & --                    & --   \\
            \hline
            SAD with auxiliary tasks and other-play~\cite{otherplay} &   22.07 (0.11) & \cellcolor{green!50}$\approx$ 0 & --                 & --                            & --                    & --                            & --                    & --   \\
            \hline
        \end{tabular}\label{stats_sig}
    \end{table}

    These results demonstrate that majority of the Rainbow with \emph{conventions} agents' performances are statistically different from existing research conducted on the Hanabi
    problem for self-play as well as cross-play agents.  The Rainbow with \emph{conventions} agent's scores are statistically similar to that of baseline Rainbow
    \revision{two-players} and ACHA \revision{three-players}, however \emph{conventions} allow the Rainbow algorithm to reach these scores significantly faster (5x compared to
    baseline Rainbow and 1000x compared to ACHA). Furthermore, even though these two instances are statistically similar, the agents learn completely different policies, as seen by
    the cross-play performance and its statistical differences. 

\end{appendices}

% \newpage
\bibliography{brede_refs.bib}% common bib file


\begin{thebibliography}{63}
\ifx \bisbn   \undefined \def \bisbn  #1{ISBN #1}\fi
\ifx \binits  \undefined \def \binits#1{#1}\fi
\ifx \bauthor  \undefined \def \bauthor#1{#1}\fi
\ifx \batitle  \undefined \def \batitle#1{#1}\fi
\ifx \bjtitle  \undefined \def \bjtitle#1{#1}\fi
\ifx \bvolume  \undefined \def \bvolume#1{\textbf{#1}}\fi
\ifx \byear  \undefined \def \byear#1{#1}\fi
\ifx \bissue  \undefined \def \bissue#1{#1}\fi
\ifx \bfpage  \undefined \def \bfpage#1{#1}\fi
\ifx \blpage  \undefined \def \blpage #1{#1}\fi
\ifx \burl  \undefined \def \burl#1{\textsf{#1}}\fi
\ifx \doiurl  \undefined \def \doiurl#1{\url{https://doi.org/#1}}\fi
\ifx \betal  \undefined \def \betal{\textit{et al.}}\fi
\ifx \binstitute  \undefined \def \binstitute#1{#1}\fi
\ifx \binstitutionaled  \undefined \def \binstitutionaled#1{#1}\fi
\ifx \bctitle  \undefined \def \bctitle#1{#1}\fi
\ifx \beditor  \undefined \def \beditor#1{#1}\fi
\ifx \bpublisher  \undefined \def \bpublisher#1{#1}\fi
\ifx \bbtitle  \undefined \def \bbtitle#1{#1}\fi
\ifx \bedition  \undefined \def \bedition#1{#1}\fi
\ifx \bseriesno  \undefined \def \bseriesno#1{#1}\fi
\ifx \blocation  \undefined \def \blocation#1{#1}\fi
\ifx \bsertitle  \undefined \def \bsertitle#1{#1}\fi
\ifx \bsnm \undefined \def \bsnm#1{#1}\fi
\ifx \bsuffix \undefined \def \bsuffix#1{#1}\fi
\ifx \bparticle \undefined \def \bparticle#1{#1}\fi
\ifx \barticle \undefined \def \barticle#1{#1}\fi
\bibcommenthead
\ifx \bconfdate \undefined \def \bconfdate #1{#1}\fi
\ifx \botherref \undefined \def \botherref #1{#1}\fi
\ifx \url \undefined \def \url#1{\textsf{#1}}\fi
\ifx \bchapter \undefined \def \bchapter#1{#1}\fi
\ifx \bbook \undefined \def \bbook#1{#1}\fi
\ifx \bcomment \undefined \def \bcomment#1{#1}\fi
\ifx \oauthor \undefined \def \oauthor#1{#1}\fi
\ifx \citeauthoryear \undefined \def \citeauthoryear#1{#1}\fi
\ifx \endbibitem  \undefined \def \endbibitem {}\fi
\ifx \bconflocation  \undefined \def \bconflocation#1{#1}\fi
\ifx \arxivurl  \undefined \def \arxivurl#1{\textsf{#1}}\fi
\csname PreBibitemsHook\endcsname

%%% 1
\bibitem[\protect\citeauthoryear{Bu{\c{s}}oniu et~al.}{2010}]{marl_why_good}
\begin{bbook}
\bauthor{\bsnm{Bu{\c{s}}oniu}, \binits{L.}},
\bauthor{\bsnm{Babu{\v{s}}ka}, \binits{R.}},
\bauthor{\bsnm{De~Schutter}, \binits{B.}}:
In: \beditor{\bsnm{Srinivasan}, \binits{D.}},
\beditor{\bsnm{Jain}, \binits{L.C.}} (eds.)
\bbtitle{Multi-agent Reinforcement Learning: An Overview},
pp. \bfpage{183}--\blpage{221}.
\bpublisher{Springer},
\blocation{Berlin, Heidelberg}
(\byear{2010}).
\doiurl{10.1007/978-3-642-14435-6_7}
\end{bbook}
\endbibitem

%%% 2
\bibitem[\protect\citeauthoryear{Cao et~al.}{2013}]{automated_MARL_vehicles}
\begin{barticle}
\bauthor{\bsnm{Cao}, \binits{Y.}},
\bauthor{\bsnm{Yu}, \binits{W.}},
\bauthor{\bsnm{Ren}, \binits{W.}},
\bauthor{\bsnm{Chen}, \binits{G.}}:
\batitle{An overview of recent progress in the study of distributed multi-agent coordination}.
\bjtitle{IEEE Transactions on Industrial Informatics}
\bvolume{9}(\bissue{1}),
\bfpage{427}--\blpage{438}
(\byear{2013})
\doiurl{10.1109/TII.2012.2219061}
\end{barticle}
\endbibitem

%%% 3
\bibitem[\protect\citeauthoryear{Hüttenrauch et~al.}{2017}]{guided_drone_swarms}
\begin{barticle}
\bauthor{\bsnm{Hüttenrauch}, \binits{M.}},
\bauthor{\bsnm{Šošić}, \binits{A.}},
\bauthor{\bsnm{Neumann}, \binits{G.}}:
\batitle{Guided deep reinforcement learning for swarm systems}.
\bjtitle{arXiv preprint arXiv:1709.06011}
(\byear{2017})
\doiurl{10.48550/arXiv.1709.06011}
\end{barticle}
\endbibitem

%%% 4
\bibitem[\protect\citeauthoryear{Branavan et~al.}{2009}]{map_instr_to_act}
\begin{bchapter}
\bauthor{\bsnm{Branavan}, \binits{S.R.}},
\bauthor{\bsnm{Chen}, \binits{H.}},
\bauthor{\bsnm{Zettlemoyer}, \binits{L.}},
\bauthor{\bsnm{Barzilay}, \binits{R.}}:
\bctitle{Reinforcement learning for mapping instructions to actions}.
In: \bbtitle{Proceedings of the Joint Conference of the 47th Annual Meeting of the ACL and the 4th International Joint Conference on Natural Language Processing of the AFNLP},
pp. \bfpage{82}--\blpage{90}
(\byear{2009})
\end{bchapter}
\endbibitem

%%% 5
\bibitem[\protect\citeauthoryear{Busoniu et~al.}{2008}]{MARL_survey}
\begin{barticle}
\bauthor{\bsnm{Busoniu}, \binits{L.}},
\bauthor{\bsnm{Babuska}, \binits{R.}},
\bauthor{\bsnm{De~Schutter}, \binits{B.}}:
\batitle{A comprehensive survey of multiagent reinforcement learning}.
\bjtitle{IEEE Transactions on Systems, Man, and Cybernetics, Part C (Applications and Reviews)}
\bvolume{38}(\bissue{2}),
\bfpage{156}--\blpage{172}
(\byear{2008})
\doiurl{10.1109/TSMCC.2007.913919}
\end{barticle}
\endbibitem

%%% 6
\bibitem[\protect\citeauthoryear{Tampuu et~al.}{2017}]{idqn_pong}
\begin{barticle}
\bauthor{\bsnm{Tampuu}, \binits{A.}},
\bauthor{\bsnm{Matiisen}, \binits{T.}},
\bauthor{\bsnm{Kodelja}, \binits{D.}},
\bauthor{\bsnm{Kuzovkin}, \binits{I.}},
\bauthor{\bsnm{Korjus}, \binits{K.}},
\bauthor{\bsnm{Aru}, \binits{J.}},
\bauthor{\bsnm{Aru}, \binits{J.}},
\bauthor{\bsnm{Vicente}, \binits{R.}}:
\batitle{Multiagent cooperation and competition with deep reinforcement learning}.
\bjtitle{PloS one}
\bvolume{12}(\bissue{4}),
\bfpage{0172395}
(\byear{2017})
\doiurl{10.1371/journal.pone.0172395}
\end{barticle}
\endbibitem

%%% 7
\bibitem[\protect\citeauthoryear{Claus and Boutilier}{1998}]{iql_short_og}
\begin{barticle}
\bauthor{\bsnm{Claus}, \binits{C.}},
\bauthor{\bsnm{Boutilier}, \binits{C.}}:
\batitle{The dynamics of reinforcement learning in cooperative multiagent systems}.
\bjtitle{AAAI/IAAI}
\bvolume{1998}(\bissue{746-752}),
\bfpage{2}
(\byear{1998})
\end{barticle}
\endbibitem

%%% 8
\bibitem[\protect\citeauthoryear{Matignon et~al.}{2012}]{iql_short_new}
\begin{barticle}
\bauthor{\bsnm{Matignon}, \binits{L.}},
\bauthor{\bsnm{Laurent}, \binits{G.J.}},
\bauthor{\bsnm{Le~Fort-Piat}, \binits{N.}}:
\batitle{Independent reinforcement learners in cooperative {Markov} games: a survey regarding coordination problems}.
\bjtitle{The Knowledge Engineering Review}
\bvolume{27}(\bissue{1}),
\bfpage{1}--\blpage{31}
(\byear{2012})
\doiurl{10.1017/S0269888912000057}
\end{barticle}
\endbibitem

%%% 9
\bibitem[\protect\citeauthoryear{Hausknecht and Stone}{2015}]{drqn_stone}
\begin{bchapter}
\bauthor{\bsnm{Hausknecht}, \binits{M.J.}},
\bauthor{\bsnm{Stone}, \binits{P.}}:
\bctitle{Deep recurrent {Q-Learning} for partially observable {MDPs}.}
In: \bbtitle{AAAI Fall Symposia},
vol. \bseriesno{45},
p. \bfpage{141}
(\byear{2015})
\end{bchapter}
\endbibitem

%%% 10
\bibitem[\protect\citeauthoryear{Sunehag et~al.}{2017}]{vdns}
\begin{barticle}
\bauthor{\bsnm{Sunehag}, \binits{P.}},
\bauthor{\bsnm{Lever}, \binits{G.}},
\bauthor{\bsnm{Gruslys}, \binits{A.}},
\bauthor{\bsnm{Czarnecki}, \binits{W.M.}},
\bauthor{\bsnm{Zambaldi}, \binits{V.}},
\bauthor{\bsnm{Jaderberg}, \binits{M.}},
\bauthor{\bsnm{Lanctot}, \binits{M.}},
\bauthor{\bsnm{Sonnerat}, \binits{N.}},
\bauthor{\bsnm{Leibo}, \binits{J.Z.}},
\bauthor{\bsnm{Tuyls}, \binits{K.}},
\bauthor{\bsnm{Graepel}, \binits{T.}}:
\batitle{Value-decomposition networks for cooperative multi-agent learning}.
\bjtitle{arXiv preprint arXiv:1706.05296}
(\byear{2017})
\doiurl{10.48550/arXiv.1706.05296}
\end{barticle}
\endbibitem

%%% 11
\bibitem[\protect\citeauthoryear{Rashid et~al.}{2020}]{rashid2018qmix}
\begin{barticle}
\bauthor{\bsnm{Rashid}, \binits{T.}},
\bauthor{\bsnm{Samvelyan}, \binits{M.}},
\bauthor{\bsnm{Witt}, \binits{C.S.}},
\bauthor{\bsnm{Farquhar}, \binits{G.}},
\bauthor{\bsnm{Foerster}, \binits{J.}},
\bauthor{\bsnm{Whiteson}, \binits{S.}}:
\batitle{Monotonic value function factorisation for deep multi-agent reinforcement learning}.
\bjtitle{Journal of Machine Learning Research}
\bvolume{21}(\bissue{178}),
\bfpage{1}--\blpage{51}
(\byear{2020})
\end{barticle}
\endbibitem

%%% 12
\bibitem[\protect\citeauthoryear{Foerster et~al.}{2016}]{foerster_rial_dial}
\begin{bchapter}
\bauthor{\bsnm{Foerster}, \binits{J.}},
\bauthor{\bsnm{Assael}, \binits{I.A.}},
\bauthor{\bsnm{Freitas}, \binits{N.}},
\bauthor{\bsnm{Whiteson}, \binits{S.}}:
\bctitle{Learning to communicate with deep multi-agent reinforcement learning}.
In: \bbtitle{Advances in Neural Information Processing Systems},
vol. \bseriesno{29}
(\byear{2016}).
\burl{https://proceedings.neurips.cc/paper_files/paper/2016/file/c7635bfd99248a2cdef8249ef7bfbef4-Paper.pdf}
\end{bchapter}
\endbibitem

%%% 13
\bibitem[\protect\citeauthoryear{Hu and Foerster}{2020}]{sad_hanabi}
\begin{bchapter}
\bauthor{\bsnm{Hu}, \binits{H.}},
\bauthor{\bsnm{Foerster}, \binits{J.N.}}:
\bctitle{Simplified action decoder for deep multi-agent reinforcement learning}.
In: \bbtitle{International Conference on Learning Representations}
(\byear{2020})
\end{bchapter}
\endbibitem

%%% 14
\bibitem[\protect\citeauthoryear{Lerer et~al.}{2020}]{sad_search}
\begin{bchapter}
\bauthor{\bsnm{Lerer}, \binits{A.}},
\bauthor{\bsnm{Hu}, \binits{H.}},
\bauthor{\bsnm{Foerster}, \binits{J.}},
\bauthor{\bsnm{Brown}, \binits{N.}}:
\bctitle{Improving policies via search in cooperative partially observable games}.
In: \bbtitle{Proceedings of the AAAI Conference on Artificial Intelligence},
vol. \bseriesno{34},
pp. \bfpage{7187}--\blpage{7194}
(\byear{2020}).
\doiurl{10.1609/aaai.v34i05.6208}
\end{bchapter}
\endbibitem

%%% 15
\bibitem[\protect\citeauthoryear{Ng et~al.}{1999}]{og_reward_shaping_1}
\begin{bchapter}
\bauthor{\bsnm{Ng}, \binits{A.Y.}},
\bauthor{\bsnm{Harada}, \binits{D.}},
\bauthor{\bsnm{Russell}, \binits{S.}}:
\bctitle{Policy invariance under reward transformations: Theory and application to reward shaping}.
In: \bbtitle{Icml},
vol. \bseriesno{99},
pp. \bfpage{278}--\blpage{287}
(\byear{1999})
\end{bchapter}
\endbibitem

%%% 16
\bibitem[\protect\citeauthoryear{Randl{\o}v and Alstr{\o}m}{1998}]{og_reward_shaping_2}
\begin{bchapter}
\bauthor{\bsnm{Randl{\o}v}, \binits{J.}},
\bauthor{\bsnm{Alstr{\o}m}, \binits{P.}}:
\bctitle{Learning to drive a bicycle using reinforcement learning and shaping.}
In: \bbtitle{ICML},
vol. \bseriesno{98},
pp. \bfpage{463}--\blpage{471}
(\byear{1998})
\end{bchapter}
\endbibitem

%%% 17
\bibitem[\protect\citeauthoryear{Devlin and Kudenko}{2011}]{marl_reward_shaping}
\begin{bchapter}
\bauthor{\bsnm{Devlin}, \binits{S.}},
\bauthor{\bsnm{Kudenko}, \binits{D.}}:
\bctitle{Theoretical considerations of potential-based reward shaping for multi-agent systems}.
In: \bbtitle{Tenth International Conference on Autonomous Agents and Multi-Agent Systems},
pp. \bfpage{225}--\blpage{232}
(\byear{2011}).
\bcomment{ACM}
\end{bchapter}
\endbibitem

%%% 18
\bibitem[\protect\citeauthoryear{Jaderberg et~al.}{2016}]{aux_tasks}
\begin{barticle}
\bauthor{\bsnm{Jaderberg}, \binits{M.}},
\bauthor{\bsnm{Mnih}, \binits{V.}},
\bauthor{\bsnm{Czarnecki}, \binits{W.M.}},
\bauthor{\bsnm{Schaul}, \binits{T.}},
\bauthor{\bsnm{Leibo}, \binits{J.Z.}},
\bauthor{\bsnm{Silver}, \binits{D.}},
\bauthor{\bsnm{Kavukcuoglu}, \binits{K.}}:
\batitle{Reinforcement learning with unsupervised auxiliary tasks}.
\bjtitle{arXiv preprint arXiv:1611.05397}
(\byear{2016})
\doiurl{10.48550/arXiv.1611.05397}
\end{barticle}
\endbibitem

%%% 19
\bibitem[\protect\citeauthoryear{Mirowski et~al.}{2016}]{aux_tasks_2}
\begin{barticle}
\bauthor{\bsnm{Mirowski}, \binits{P.}},
\bauthor{\bsnm{Pascanu}, \binits{R.}},
\bauthor{\bsnm{Viola}, \binits{F.}},
\bauthor{\bsnm{Soyer}, \binits{H.}},
\bauthor{\bsnm{Ballard}, \binits{A.J.}},
\bauthor{\bsnm{Banino}, \binits{A.}},
\bauthor{\bsnm{Denil}, \binits{M.}},
\bauthor{\bsnm{Goroshin}, \binits{R.}},
\bauthor{\bsnm{Sifre}, \binits{L.}},
\bauthor{\bsnm{Kavukcuoglu}, \binits{K.}}, \betal:
\batitle{Learning to navigate in complex environments}.
\bjtitle{arXiv preprint arXiv:1611.03673}
(\byear{2016})
\doiurl{10.48550/arXiv.1611.03673}
\end{barticle}
\endbibitem

%%% 20
\bibitem[\protect\citeauthoryear{Sutton et~al.}{1999}]{sutton1999options}
\begin{barticle}
\bauthor{\bsnm{Sutton}, \binits{R.S.}},
\bauthor{\bsnm{Precup}, \binits{D.}},
\bauthor{\bsnm{Singh}, \binits{S.}}:
\batitle{Between mdps and semi-mdps: A framework for temporal abstraction in reinforcement learning}.
\bjtitle{Artificial Intelligence}
\bvolume{112}(\bissue{1}),
\bfpage{181}--\blpage{211}
(\byear{1999})
\doiurl{10.1016/S0004-3702(99)00052-1}
\end{barticle}
\endbibitem

%%% 21
\bibitem[\protect\citeauthoryear{Silver et~al.}{2016}]{silver2016alphago}
\begin{barticle}
\bauthor{\bsnm{Silver}, \binits{D.}},
\bauthor{\bsnm{Huang}, \binits{A.}},
\bauthor{\bsnm{Maddison}, \binits{C.J.}},
\bauthor{\bsnm{Guez}, \binits{A.}},
\bauthor{\bsnm{Sifre}, \binits{L.}},
\bauthor{\bsnm{Van Den~Driessche}, \binits{G.}},
\bauthor{\bsnm{Schrittwieser}, \binits{J.}},
\bauthor{\bsnm{Antonoglou}, \binits{I.}},
\bauthor{\bsnm{Panneershelvam}, \binits{V.}},
\bauthor{\bsnm{Lanctot}, \binits{M.}}, \betal:
\batitle{Mastering the game of {Go} with deep neural networks and tree search}.
\bjtitle{Nature}
\bvolume{529}(\bissue{7587}),
\bfpage{484}--\blpage{489}
(\byear{2016})
\end{barticle}
\endbibitem

%%% 22
\bibitem[\protect\citeauthoryear{Tesauro}{1995}]{gammon}
\begin{barticle}
\bauthor{\bsnm{Tesauro}, \binits{G.}}:
\batitle{Temporal difference learning and {TD-Gammon}}.
\bjtitle{Communications of the ACM}
\bvolume{38}(\bissue{3}),
\bfpage{58}--\blpage{68}
(\byear{1995})
\end{barticle}
\endbibitem

%%% 23
\bibitem[\protect\citeauthoryear{Berner et~al.}{2019}]{openai_five}
\begin{barticle}
\bauthor{\bsnm{Berner}, \binits{C.}},
\bauthor{\bsnm{Brockman}, \binits{G.}},
\bauthor{\bsnm{Chan}, \binits{B.}},
\bauthor{\bsnm{Cheung}, \binits{V.}},
\bauthor{\bsnm{Debiak}, \binits{P.}},
\bauthor{\bsnm{Dennison}, \binits{C.}},
\bauthor{\bsnm{Farhi}, \binits{D.}},
\bauthor{\bsnm{Fischer}, \binits{Q.}},
\bauthor{\bsnm{Hashme}, \binits{S.}},
\bauthor{\bsnm{Hesse}, \binits{C.}},
\bauthor{\bsnm{Józefowicz}, \binits{R.}},
\bauthor{\bsnm{Gray}, \binits{S.}},
\bauthor{\bsnm{Olsson}, \binits{C.}},
\bauthor{\bsnm{Pachocki}, \binits{J.}},
\bauthor{\bsnm{Petrov}, \binits{M.}},
\bauthor{\bsnm{Pinto}, \binits{H.P.d.O.}},
\bauthor{\bsnm{Raiman}, \binits{J.}},
\bauthor{\bsnm{Salimans}, \binits{T.}},
\bauthor{\bsnm{Schlatter}, \binits{J.}},
\bauthor{\bsnm{Schneider}, \binits{J.}},
\bauthor{\bsnm{Sidor}, \binits{S.}},
\bauthor{\bsnm{Sutskever}, \binits{I.}},
\bauthor{\bsnm{Tang}, \binits{J.}},
\bauthor{\bsnm{Wolski}, \binits{F.}},
\bauthor{\bsnm{Zhang}, \binits{S.}}:
\batitle{Dota 2 with large scale deep reinforcement learning}.
\bjtitle{arXiv preprint arXiv:1912.06680}
(\byear{2019})
\doiurl{10.48550/arXiv.1912.06680}
\end{barticle}
\endbibitem

%%% 24
\bibitem[\protect\citeauthoryear{Premack and Woodruff}{1978}]{ToM}
\begin{botherref}
\oauthor{\bsnm{Premack}, \binits{D.}},
\oauthor{\bsnm{Woodruff}, \binits{G.}}:
Does the chimpanzee have a theory of mind?
Behavioral and Brain Science 1,
515--526
(1978)
\end{botherref}
\endbibitem

%%% 25
\bibitem[\protect\citeauthoryear{Lewis}{2008}]{lewis2008convention}
\begin{bbook}
\bauthor{\bsnm{Lewis}, \binits{D.}}:
\bbtitle{Convention: A Philosophical Study},
\bedition{1}st edn.
\bpublisher{Blackwell Publishers Ltd},
\blocation{Hoboken, New Jersey}
(\byear{2008})
\end{bbook}
\endbibitem

%%% 26
\bibitem[\protect\citeauthoryear{Foerster et~al.}{2019}]{bad}
\begin{bchapter}
\bauthor{\bsnm{Foerster}, \binits{J.}},
\bauthor{\bsnm{Song}, \binits{F.}},
\bauthor{\bsnm{Hughes}, \binits{E.}},
\bauthor{\bsnm{Burch}, \binits{N.}},
\bauthor{\bsnm{Dunning}, \binits{I.}},
\bauthor{\bsnm{Whiteson}, \binits{S.}},
\bauthor{\bsnm{Botvinick}, \binits{M.}},
\bauthor{\bsnm{Bowling}, \binits{M.}}:
\bctitle{Bayesian action decoder for deep multi-agent reinforcement learning}.
In: \bbtitle{International Conference on Machine Learning},
pp. \bfpage{1942}--\blpage{1951}
(\byear{2019}).
\bcomment{PMLR}
\end{bchapter}
\endbibitem

%%% 27
\bibitem[\protect\citeauthoryear{Silver et~al.}{2017}]{self_play}
\begin{barticle}
\bauthor{\bsnm{Silver}, \binits{D.}},
\bauthor{\bsnm{Hubert}, \binits{T.}},
\bauthor{\bsnm{Schrittwieser}, \binits{J.}},
\bauthor{\bsnm{Antonoglou}, \binits{I.}},
\bauthor{\bsnm{Lai}, \binits{M.}},
\bauthor{\bsnm{Guez}, \binits{A.}},
\bauthor{\bsnm{Lanctot}, \binits{M.}},
\bauthor{\bsnm{Sifre}, \binits{L.}},
\bauthor{\bsnm{Kumaran}, \binits{D.}},
\bauthor{\bsnm{Graepel}, \binits{T.}},
\bauthor{\bsnm{Lillicrap}, \binits{T.}},
\bauthor{\bsnm{Simonyan}, \binits{K.}},
\bauthor{\bsnm{Hassabis}, \binits{D.}}:
\batitle{Mastering {Chess} and {Shogi} by {Self-Play} with a {General Reinforcement Learning Algorithm}}.
\bjtitle{arXiv preprint arXiv:1712.01815}
(\byear{2017})
\doiurl{10.48550/arXiv.1712.01815}
\end{barticle}
\endbibitem

%%% 28
\bibitem[\protect\citeauthoryear{Hu et~al.}{2020}]{otherplay}
\begin{bchapter}
\bauthor{\bsnm{Hu}, \binits{H.}},
\bauthor{\bsnm{Lerer}, \binits{A.}},
\bauthor{\bsnm{Peysakhovich}, \binits{A.}},
\bauthor{\bsnm{Foerster}, \binits{J.}}:
\bctitle{{"Other-Play"} for zero-shot coordination}.
In: \bbtitle{International Conference on Machine Learning},
pp. \bfpage{4399}--\blpage{4410}
(\byear{2020}).
\bcomment{PMLR}
\end{bchapter}
\endbibitem

%%% 29
\bibitem[\protect\citeauthoryear{Hessel et~al.}{2018}]{rainbow}
\begin{bchapter}
\bauthor{\bsnm{Hessel}, \binits{M.}},
\bauthor{\bsnm{Modayil}, \binits{J.}},
\bauthor{\bsnm{Van~Hasselt}, \binits{H.}},
\bauthor{\bsnm{Schaul}, \binits{T.}},
\bauthor{\bsnm{Ostrovski}, \binits{G.}},
\bauthor{\bsnm{Dabney}, \binits{W.}},
\bauthor{\bsnm{Horgan}, \binits{D.}},
\bauthor{\bsnm{Piot}, \binits{B.}},
\bauthor{\bsnm{Azar}, \binits{M.}},
\bauthor{\bsnm{Silver}, \binits{D.}}:
\bctitle{Rainbow: Combining improvements in deep reinforcement learning}.
In: \bbtitle{Thirty-second AAAI Conference on Artificial Intelligence}
(\byear{2018}).
\doiurl{10.1609/aaai.v32i1.11796}
\end{bchapter}
\endbibitem

%%% 30
\bibitem[\protect\citeauthoryear{Harrold et~al.}{2022}]{rainbow_energy}
\begin{barticle}
\bauthor{\bsnm{Harrold}, \binits{D.J.B.}},
\bauthor{\bsnm{Cao}, \binits{J.}},
\bauthor{\bsnm{Fan}, \binits{Z.}}:
\batitle{Data-driven battery operation for energy arbitrage using rainbow deep reinforcement learning}.
\bjtitle{Energy}
\bvolume{238},
\bfpage{121958}
(\byear{2022})
\doiurl{10.1016/j.energy.2021.121958}
\end{barticle}
\endbibitem

%%% 31
\bibitem[\protect\citeauthoryear{Yang et~al.}{2020}]{rainbow_good_1}
\begin{barticle}
\bauthor{\bsnm{Yang}, \binits{J.}},
\bauthor{\bsnm{Yang}, \binits{M.}},
\bauthor{\bsnm{Wang}, \binits{M.}},
\bauthor{\bsnm{Du}, \binits{P.}},
\bauthor{\bsnm{Yu}, \binits{Y.}}:
\batitle{A deep reinforcement learning method for managing wind farm uncertainties through energy storage system control and external reserve purchasing}.
\bjtitle{International Journal of Electrical Power \& Energy Systems}
\bvolume{119},
\bfpage{105928}
(\byear{2020})
\doiurl{10.1016/j.ijepes.2020.105928}
\end{barticle}
\endbibitem

%%% 32
\bibitem[\protect\citeauthoryear{Wang et~al.}{2022}]{rainbow_good_2}
\begin{barticle}
\bauthor{\bsnm{Wang}, \binits{R.}},
\bauthor{\bsnm{Chen}, \binits{Z.}},
\bauthor{\bsnm{Xing}, \binits{Q.}},
\bauthor{\bsnm{Zhang}, \binits{Z.}},
\bauthor{\bsnm{Zhang}, \binits{T.}}:
\batitle{A modified rainbow-based deep reinforcement learning method for optimal scheduling of charging station}.
\bjtitle{Sustainability}
\bvolume{14}(\bissue{3}),
\bfpage{1884}
(\byear{2022})
\doiurl{10.3390/su14031884}
\end{barticle}
\endbibitem

%%% 33
\bibitem[\protect\citeauthoryear{Harrold et~al.}{2022}]{marainbow}
\begin{barticle}
\bauthor{\bsnm{Harrold}, \binits{D.J.B.}},
\bauthor{\bsnm{Cao}, \binits{J.}},
\bauthor{\bsnm{Fan}, \binits{Z.}}:
\batitle{Renewable energy integration and microgrid energy trading using multi-agent deep reinforcement learning}.
\bjtitle{Applied Energy}
\bvolume{318},
\bfpage{119151}
(\byear{2022})
\doiurl{10.1016/j.apenergy.2022.119151}
\end{barticle}
\endbibitem

%%% 34
\bibitem[\protect\citeauthoryear{Zhao et~al.}{2019}]{rainbow_shortcoming}
\begin{barticle}
\bauthor{\bsnm{Zhao}, \binits{Y.}},
\bauthor{\bsnm{Borovikov}, \binits{I.}},
\bauthor{\bsnm{Rupert}, \binits{J.}},
\bauthor{\bsnm{Somers}, \binits{C.}},
\bauthor{\bsnm{Beirami}, \binits{A.}}:
\batitle{On multi-agent learning in team sports games}.
\bjtitle{arXiv preprint arXiv:1906.10124}
(\byear{2019})
\doiurl{10.48550/ARXIV.1906.10124}
\end{barticle}
\endbibitem

%%% 35
\bibitem[\protect\citeauthoryear{Ceron and Castro}{2021}]{rainbow_suboptimal_1}
\begin{bchapter}
\bauthor{\bsnm{Ceron}, \binits{J.S.O.}},
\bauthor{\bsnm{Castro}, \binits{P.S.}}:
\bctitle{Revisiting rainbow: Promoting more insightful and inclusive deep reinforcement learning research}.
In: \bbtitle{International Conference on Machine Learning},
pp. \bfpage{1373}--\blpage{1383}
(\byear{2021}).
\bcomment{PMLR}
\end{bchapter}
\endbibitem

%%% 36
\bibitem[\protect\citeauthoryear{Toromanoff et~al.}{2019}]{rainbow_suboptimal_2}
\begin{barticle}
\bauthor{\bsnm{Toromanoff}, \binits{M.}},
\bauthor{\bsnm{Wirbel}, \binits{E.}},
\bauthor{\bsnm{Moutarde}, \binits{F.}}:
\batitle{Is deep reinforcement learning really superhuman on atari? leveling the playing field}.
\bjtitle{arXiv preprint arXiv:1908.04683}
(\byear{2019})
\doiurl{10.48550/arXiv.1908.04683}
\end{barticle}
\endbibitem

%%% 37
\bibitem[\protect\citeauthoryear{Bard et~al.}{2020}]{hanabi_ai}
\begin{barticle}
\bauthor{\bsnm{Bard}, \binits{N.}},
\bauthor{\bsnm{Foerster}, \binits{J.N.}},
\bauthor{\bsnm{Chandar}, \binits{S.}},
\bauthor{\bsnm{Burch}, \binits{N.}},
\bauthor{\bsnm{Lanctot}, \binits{M.}},
\bauthor{\bsnm{Song}, \binits{H.F.}},
\bauthor{\bsnm{Parisotto}, \binits{E.}},
\bauthor{\bsnm{Dumoulin}, \binits{V.}},
\bauthor{\bsnm{Moitra}, \binits{S.}},
\bauthor{\bsnm{Hughes}, \binits{E.}},
\bauthor{\bsnm{Dunning}, \binits{I.}},
\bauthor{\bsnm{Mourad}, \binits{S.}},
\bauthor{\bsnm{Larochelle}, \binits{H.}},
\bauthor{\bsnm{Bellemare}, \binits{M.G.}},
\bauthor{\bsnm{Bowling}, \binits{M.}}:
\batitle{The {Hanabi} challenge: A new frontier for {AI} research}.
\bjtitle{Artificial Intelligence}
\bvolume{280},
\bfpage{103216}
(\byear{2020})
\doiurl{10.1016/j.artint.2019.103216}
\end{barticle}
\endbibitem

%%% 38
\bibitem[\protect\citeauthoryear{Cox et~al.}{2015}]{hatbot}
\begin{barticle}
\bauthor{\bsnm{Cox}, \binits{C.}},
\bauthor{\bsnm{De~Silva}, \binits{J.}},
\bauthor{\bsnm{Deorsey}, \binits{P.}},
\bauthor{\bsnm{Kenter}, \binits{F.H.}},
\bauthor{\bsnm{Retter}, \binits{T.}},
\bauthor{\bsnm{Tobin}, \binits{J.}}:
\batitle{How to make the perfect fireworks display: Two strategies for hanabi}.
\bjtitle{Mathematics Magazine}
\bvolume{88}(\bissue{5}),
\bfpage{323}--\blpage{336}
(\byear{2015})
\end{barticle}
\endbibitem

%%% 39
\bibitem[\protect\citeauthoryear{Wu}{}]{wtfwthat}
\begin{botherref}
\oauthor{\bsnm{Wu}, \binits{J.}}:
Github - wuthefwasthat/hanabi.rs: Hanabi simulation in rust.
[Online; accessed 17 July 2024].
\url{https://github.com/WuTheFWasThat/hanabi.rs}
\end{botherref}
\endbibitem

%%% 40
\bibitem[\protect\citeauthoryear{Brown and Sandholm}{2019}]{multi_agent_search}
\begin{barticle}
\bauthor{\bsnm{Brown}, \binits{N.}},
\bauthor{\bsnm{Sandholm}, \binits{T.}}:
\batitle{Superhuman ai for multiplayer poker}.
\bjtitle{Science}
\bvolume{365}(\bissue{6456}),
\bfpage{885}--\blpage{890}
(\byear{2019})
\doiurl{10.1126/science.aay2400}
\end{barticle}
\endbibitem

%%% 41
\bibitem[\protect\citeauthoryear{Yu et~al.}{2022}]{mappo}
\begin{barticle}
\bauthor{\bsnm{Yu}, \binits{C.}},
\bauthor{\bsnm{Velu}, \binits{A.}},
\bauthor{\bsnm{Vinitsky}, \binits{E.}},
\bauthor{\bsnm{Gao}, \binits{J.}},
\bauthor{\bsnm{Wang}, \binits{Y.}},
\bauthor{\bsnm{Bayen}, \binits{A.}},
\bauthor{\bsnm{WU}, \binits{Y.}}:
\batitle{The surprising effectiveness of ppo in cooperative multi-agent games}.
\bjtitle{Advances in neural information processing systems}
\bvolume{35},
\bfpage{24611}--\blpage{24624}
(\byear{2022})
\end{barticle}
\endbibitem

%%% 42
\bibitem[\protect\citeauthoryear{Lowe et~al.}{2017}]{mpe}
\begin{botherref}
\oauthor{\bsnm{Lowe}, \binits{R.}},
\oauthor{\bsnm{Wu}, \binits{Y.I.}},
\oauthor{\bsnm{Tamar}, \binits{A.}},
\oauthor{\bsnm{Harb}, \binits{J.}},
\oauthor{\bsnm{Pieter~Abbeel}, \binits{O.}},
\oauthor{\bsnm{Mordatch}, \binits{I.}}:
Multi-agent actor-critic for mixed cooperative-competitive environments.
Advances in neural information processing systems
\textbf{30}
(2017)
\end{botherref}
\endbibitem

%%% 43
\bibitem[\protect\citeauthoryear{Vinyals et~al.}{2019}]{smac}
\begin{barticle}
\bauthor{\bsnm{Vinyals}, \binits{O.}},
\bauthor{\bsnm{Babuschkin}, \binits{I.}},
\bauthor{\bsnm{Czarnecki}, \binits{W.M.}},
\bauthor{\bsnm{Mathieu}, \binits{M.}},
\bauthor{\bsnm{Dudzik}, \binits{A.}},
\bauthor{\bsnm{Chung}, \binits{J.}},
\bauthor{\bsnm{Choi}, \binits{D.H.}},
\bauthor{\bsnm{Powell}, \binits{R.}},
\bauthor{\bsnm{Ewalds}, \binits{T.}},
\bauthor{\bsnm{Georgiev}, \binits{P.}}, \betal:
\batitle{Grandmaster level in starcraft ii using multi-agent reinforcement learning}.
\bjtitle{Nature}
\bvolume{575}(\bissue{7782}),
\bfpage{350}--\blpage{354}
(\byear{2019})
\end{barticle}
\endbibitem

%%% 44
\bibitem[\protect\citeauthoryear{Kurach et~al.}{2020}]{grf}
\begin{bchapter}
\bauthor{\bsnm{Kurach}, \binits{K.}},
\bauthor{\bsnm{Raichuk}, \binits{A.}},
\bauthor{\bsnm{Stanczyk}, \binits{P.}},
\bauthor{\bsnm{Zajac}, \binits{M.}},
\bauthor{\bsnm{Bachem}, \binits{O.}},
\bauthor{\bsnm{Espeholt}, \binits{L.}},
\bauthor{\bsnm{Riquelme}, \binits{C.}},
\bauthor{\bsnm{Vincent}, \binits{D.}},
\bauthor{\bsnm{Michalski}, \binits{M.}},
\bauthor{\bsnm{Bousquet}, \binits{O.}}, \betal:
\bctitle{Google research football: A novel reinforcement learning environment}.
In: \bbtitle{Proceedings of the AAAI Conference on Artificial Intelligence},
vol. \bseriesno{34},
pp. \bfpage{4501}--\blpage{4510}
(\byear{2020}).
\doiurl{10.1609/aaai.v34i04.5878}
\end{bchapter}
\endbibitem

%%% 45
\bibitem[\protect\citeauthoryear{Wang et~al.}{2021a}]{qplex}
\begin{bchapter}
\bauthor{\bsnm{Wang}, \binits{J.}},
\bauthor{\bsnm{Ren}, \binits{Z.}},
\bauthor{\bsnm{Liu}, \binits{T.}},
\bauthor{\bsnm{Yu}, \binits{Y.}},
\bauthor{\bsnm{Zhang}, \binits{C.}}:
\bctitle{Qplex: Duplexdueling multi-agent q-learning.}
In: \bbtitle{International Conference on Learning Representation}
(\byear{2021})
\end{bchapter}
\endbibitem

%%% 46
\bibitem[\protect\citeauthoryear{Wang et~al.}{2021b}]{rode}
\begin{bchapter}
\bauthor{\bsnm{Wang}, \binits{T.}},
\bauthor{\bsnm{Gupta}, \binits{T.}},
\bauthor{\bsnm{Mahajan}, \binits{A.}},
\bauthor{\bsnm{Peng}, \binits{B.}},
\bauthor{\bsnm{Whiteson}, \binits{S.}},
\bauthor{\bsnm{Zhang}, \binits{C.}}:
\bctitle{Rode: Learning roles to decompose multi-agent tasks}.
In: \bbtitle{International Conference on Learning Representation}
(\byear{2021})
\end{bchapter}
\endbibitem

%%% 47
\bibitem[\protect\citeauthoryear{Amato et~al.}{2019}]{macdec_pomdp}
\begin{barticle}
\bauthor{\bsnm{Amato}, \binits{C.}},
\bauthor{\bsnm{Konidaris}, \binits{G.}},
\bauthor{\bsnm{Kaelbling}, \binits{L.P.}},
\bauthor{\bsnm{How}, \binits{J.P.}}:
\batitle{Modeling and planning with macro-actions in decentralized pomdps}.
\bjtitle{Journal of Artificial Intelligence Research}
\bvolume{64},
\bfpage{817}--\blpage{859}
(\byear{2019})
\doiurl{10.1613/jair.1.11418}
\end{barticle}
\endbibitem

%%% 48
\bibitem[\protect\citeauthoryear{Chen et~al.}{2022}]{option_discovery_marl}
\begin{barticle}
\bauthor{\bsnm{Chen}, \binits{J.}},
\bauthor{\bsnm{Chen}, \binits{J.}},
\bauthor{\bsnm{Lan}, \binits{T.}},
\bauthor{\bsnm{Aggarwal}, \binits{V.}}:
\batitle{Scalable multi-agent covering option discovery based on kronecker graphs}.
\bjtitle{Advances in Neural Information Processing Systems}
\bvolume{35},
\bfpage{30406}--\blpage{30418}
(\byear{2022})
\end{barticle}
\endbibitem

%%% 49
\bibitem[\protect\citeauthoryear{Strouse et~al.}{2021}]{sp_vs_cp_1}
\begin{barticle}
\bauthor{\bsnm{Strouse}, \binits{D.}},
\bauthor{\bsnm{McKee}, \binits{K.}},
\bauthor{\bsnm{Botvinick}, \binits{M.}},
\bauthor{\bsnm{Hughes}, \binits{E.}},
\bauthor{\bsnm{Everett}, \binits{R.}}:
\batitle{Collaborating with humans without human data}.
\bjtitle{Advances in Neural Information Processing Systems}
\bvolume{34},
\bfpage{14502}--\blpage{14515}
(\byear{2021})
\end{barticle}
\endbibitem

%%% 50
\bibitem[\protect\citeauthoryear{Carroll et~al.}{2019}]{sp_vs_cp_2}
\begin{botherref}
\oauthor{\bsnm{Carroll}, \binits{M.}},
\oauthor{\bsnm{Shah}, \binits{R.}},
\oauthor{\bsnm{Ho}, \binits{M.K.}},
\oauthor{\bsnm{Griffiths}, \binits{T.}},
\oauthor{\bsnm{Seshia}, \binits{S.}},
\oauthor{\bsnm{Abbeel}, \binits{P.}},
\oauthor{\bsnm{Dragan}, \binits{A.}}:
On the utility of learning about humans for human-ai coordination.
Advances in neural information processing systems
\textbf{32}
(2019)
\end{botherref}
\endbibitem

%%% 51
\bibitem[\protect\citeauthoryear{Sutton and Barto}{2018}]{sutton_barto}
\begin{bbook}
\bauthor{\bsnm{Sutton}, \binits{R.S.}},
\bauthor{\bsnm{Barto}, \binits{A.G.}}:
\bbtitle{Reinforcement Learning: An Introduction}
vol. \bseriesno{1},
\bedition{2}nd edn.
\bpublisher{The MIT Press},
\blocation{Cambridge, Massachusetts}
(\byear{2018})
\end{bbook}
\endbibitem

%%% 52
\bibitem[\protect\citeauthoryear{Oliehoek and Amato}{2016}]{dec_pomdp}
\begin{bbook}
\bauthor{\bsnm{Oliehoek}, \binits{F.A.}},
\bauthor{\bsnm{Amato}, \binits{C.}}:
\bbtitle{A Concise Introduction to Decentralized POMDPs}
vol. \bseriesno{1},
\bedition{1}st edn.
\bpublisher{Springer},
\blocation{Cham, Switzerland}
(\byear{2016})
\end{bbook}
\endbibitem

%%% 53
\bibitem[\protect\citeauthoryear{Tan}{1993}]{iql_og}
\begin{bchapter}
\bauthor{\bsnm{Tan}, \binits{M.}}:
\bctitle{Multi-agent reinforcement learning: Independent vs. cooperative agents}.
In: \bbtitle{Proceedings of the Tenth International Conference on Machine Learning},
pp. \bfpage{330}--\blpage{337}
(\byear{1993})
\end{bchapter}
\endbibitem

%%% 54
\bibitem[\protect\citeauthoryear{Mnih et~al.}{2015}]{dqn_main}
\begin{barticle}
\bauthor{\bsnm{Mnih}, \binits{V.}},
\bauthor{\bsnm{Kavukcuoglu}, \binits{K.}},
\bauthor{\bsnm{Silver}, \binits{D.}},
\bauthor{\bsnm{Rusu}, \binits{A.A.}},
\bauthor{\bsnm{Veness}, \binits{J.}},
\bauthor{\bsnm{Bellemare}, \binits{M.G.}},
\bauthor{\bsnm{Graves}, \binits{A.}},
\bauthor{\bsnm{Riedmiller}, \binits{M.}},
\bauthor{\bsnm{Fidjeland}, \binits{A.K.}},
\bauthor{\bsnm{Ostrovski}, \binits{G.}}, \betal:
\batitle{Human-level control through deep reinforcement learning}.
\bjtitle{Nature}
\bvolume{518}(\bissue{7540}),
\bfpage{529}--\blpage{533}
(\byear{2015})
\end{barticle}
\endbibitem

%%% 55
\bibitem[\protect\citeauthoryear{Lin}{1992}]{replay_mem}
\begin{barticle}
\bauthor{\bsnm{Lin}, \binits{L.J.}}:
\batitle{Self-improving reactive agents based on reinforcement learning, planning and teaching}.
\bjtitle{Machine learning}
\bvolume{8}(\bissue{3}),
\bfpage{293}--\blpage{321}
(\byear{1992})
\doiurl{10.1007/BF00992699}
\end{barticle}
\endbibitem

%%% 56
\bibitem[\protect\citeauthoryear{Kingma and Ba}{2017}]{adam_opt}
\begin{barticle}
\bauthor{\bsnm{Kingma}, \binits{D.P.}},
\bauthor{\bsnm{Ba}, \binits{J.}}:
\batitle{Adam: A method for stochastic optimization}.
\bjtitle{arXiv preprint arXiv:1412.6980}
(\byear{2017})
\doiurl{10.48550/ARXIV.1412.6980}
\end{barticle}
\endbibitem

%%% 57
\bibitem[\protect\citeauthoryear{Machado et~al.}{2017}]{option_discovery}
\begin{bchapter}
\bauthor{\bsnm{Machado}, \binits{M.C.}},
\bauthor{\bsnm{Bellemare}, \binits{M.G.}},
\bauthor{\bsnm{Bowling}, \binits{M.}}:
\bctitle{A laplacian framework for option discovery in reinforcement learning}.
In: \bbtitle{International Conference on Machine Learning},
pp. \bfpage{2295}--\blpage{2304}
(\byear{2017}).
\bcomment{PMLR}
\end{bchapter}
\endbibitem

%%% 58
\bibitem[\protect\citeauthoryear{Schrittwieser et~al.}{2020}]{rainbow_example_1}
\begin{barticle}
\bauthor{\bsnm{Schrittwieser}, \binits{J.}},
\bauthor{\bsnm{Antonoglou}, \binits{I.}},
\bauthor{\bsnm{Hubert}, \binits{T.}},
\bauthor{\bsnm{Simonyan}, \binits{K.}},
\bauthor{\bsnm{Sifre}, \binits{L.}},
\bauthor{\bsnm{Schmitt}, \binits{S.}},
\bauthor{\bsnm{Guez}, \binits{A.}},
\bauthor{\bsnm{Lockhart}, \binits{E.}},
\bauthor{\bsnm{Hassabis}, \binits{D.}},
\bauthor{\bsnm{Graepel}, \binits{T.}}, \betal:
\batitle{Mastering atari, go, chess and shogi by planning with a learned model}.
\bjtitle{Nature}
\bvolume{588}(\bissue{7839}),
\bfpage{604}--\blpage{609}
(\byear{2020})
\end{barticle}
\endbibitem

%%% 59
\bibitem[\protect\citeauthoryear{Luong et~al.}{2019}]{rainbow_example_2}
\begin{barticle}
\bauthor{\bsnm{Luong}, \binits{N.C.}},
\bauthor{\bsnm{Hoang}, \binits{D.T.}},
\bauthor{\bsnm{Gong}, \binits{S.}},
\bauthor{\bsnm{Niyato}, \binits{D.}},
\bauthor{\bsnm{Wang}, \binits{P.}},
\bauthor{\bsnm{Liang}, \binits{Y.-C.}},
\bauthor{\bsnm{Kim}, \binits{D.I.}}:
\batitle{Applications of deep reinforcement learning in communications and networking: A survey}.
\bjtitle{IEEE communications surveys \& tutorials}
\bvolume{21}(\bissue{4}),
\bfpage{3133}--\blpage{3174}
(\byear{2019})
\doiurl{10.1109/COMST.2019.2916583}
\end{barticle}
\endbibitem

%%% 60
\bibitem[\protect\citeauthoryear{Zhang et~al.}{2019}]{rainbow_example_3}
\begin{barticle}
\bauthor{\bsnm{Zhang}, \binits{C.}},
\bauthor{\bsnm{Patras}, \binits{P.}},
\bauthor{\bsnm{Haddadi}, \binits{H.}}:
\batitle{Deep learning in mobile and wireless networking: A survey}.
\bjtitle{IEEE Communications surveys \& tutorials}
\bvolume{21}(\bissue{3}),
\bfpage{2224}--\blpage{2287}
(\byear{2019})
\doiurl{10.1109/COMST.2019.2904897}
\end{barticle}
\endbibitem

%%% 61
\bibitem[\protect\citeauthoryear{Schmidt-Hieber}{2020}]{relu}
\begin{barticle}
\bauthor{\bsnm{Schmidt-Hieber}, \binits{J.}}:
\batitle{Nonparametric regression using deep neural networks with {ReLU} activation function}.
\bjtitle{The Annals of Statistics}
\bvolume{48}(\bissue{4}),
\bfpage{1875}--\blpage{1897}
(\byear{2020})
\end{barticle}
\endbibitem

%%% 62
\bibitem[\protect\citeauthoryear{Kraemer and Banerjee}{2016}]{central_decentral}
\begin{barticle}
\bauthor{\bsnm{Kraemer}, \binits{L.}},
\bauthor{\bsnm{Banerjee}, \binits{B.}}:
\batitle{Multi-agent reinforcement learning as a rehearsal for decentralized planning}.
\bjtitle{Neurocomputing}
\bvolume{190},
\bfpage{82}--\blpage{94}
(\byear{2016})
\doiurl{10.1016/j.neucom.2016.01.031}
\end{barticle}
\endbibitem

%%% 63
\bibitem[\protect\citeauthoryear{Welch}{1947}]{welch_t_test}
\begin{barticle}
\bauthor{\bsnm{Welch}, \binits{B.L.}}:
\batitle{The generalization of ‘student's’problem when several different population varlances are involved}.
\bjtitle{Biometrika}
\bvolume{34}(\bissue{1-2}),
\bfpage{28}--\blpage{35}
(\byear{1947})
\doiurl{10.1093/biomet/34.1-2.28}
\end{barticle}
\endbibitem

\end{thebibliography}

\end{document}